\documentclass{aa} 

\usepackage{graphicx}
\usepackage{txfonts}
\usepackage{hyperref}
\usepackage{scalerel}
\usepackage{multirow}
\usepackage{amsmath}	
\usepackage{amssymb}	
\begin{document}

   \title{Identification of ram pressure stripping features in galaxies using citizen science}

    \titlerunning{Morphologies of ram pressure stripped galaxies}
   \author{
Jacob P. Crossett,\inst{1,2}\thanks{E-mail: jacob.crossett@uv.cl} 
\and
Yara L. Jaff\'{e},\inst{1,2}
\and
Sean L. McGee,\inst{3}
\and
Rory Smith,\inst{4}
\and
Callum Bellhouse,\inst{5}
\and
Daniela Bettoni,\inst{5}
\and
Benedetta Vulcani,\inst{5}
\and
Kshitija Kelkar,\inst{1,6}
\and
Ana C. C. Lourenço\inst{1,7}
}

\institute{
Departamento de Física, Universidad Técnica Federico Santa María, Avenida España 1680, Valparaíso, Chile
\and
Instituto de F\'{i}sica y Astronom\'{i}a, Universidad de Valpara\'{i}so, Avda. Gran Breta\~{n}a 1111, Valpara\'{i}so, Chile
\and
School of Physics and Astronomy, University of Birmingham, Birmingham, B15 2TT, UK
\and
Departamento de F\'{i}sicaa, Universidad T\'{e}cnica Federico Santa Mar\'{i}a, Vicu\~{n}a Mackenna 3939, San Joaqu\'{i}n, Santiago de Chile, Chile
\and
INAF-Astronomical Observatory of Padova, vicolo dell’Osservatorio 5, I-35122 Padova, Italy
\and
Vera C. Rubin Observatory, Av. Juan Cisternas 1500, La Serena, Chile
\and
European Southern Obervatory (ESO), Alonso de Cordova 3107, Santiago, Chile
}
\date{Received XXX; accepted YYY}

   \abstract
   {Ram pressure stripped galaxies are rare cases of environmental evolution in action. However, our ability to understand these transforming galaxies is limited by the small number of identified galaxies experiencing ram pressure stripping (RPS).}
   {Our aim is to explore the efficacy of citizen science classifications in identifying ram pressure stripped galaxies, and use this to aid in motivating new potential samples of ram pressure stripped candidates.}
   {We compile a sample of over 200 known ram pressure stripped galaxies from existing literature, with morphological classifications obtained from Galaxy Zoo. We compare these galaxies with magnitude and redshift-matched comparison cluster and field galaxies. Additionally, we create a sample of SDSS cluster galaxies, with morphological classifications similar to known ram pressure stripped galaxies, and compare the fraction of potential new RPS candidates against control samples.}
   {We find that ram pressure stripped galaxies exhibit a higher proportion of `odd' and `irregular' morphological classifications compared to field and cluster comparison samples. This trend is particularly pronounced in galaxies displaying strong optical ram pressure stripping features, but absent from galaxies with only radio tails. 
   We find that SDSS galaxies with Galaxy Zoo classifications consistent with the known RPS galaxies have a higher fraction of visible ram pressure stripping features ($19\%$) compared with other cluster galaxies ($12\%$) when classified by experts. We identify 101 new ram pressure stripping candidate galaxies through these expert classifications.}
  {We demonstrate that indirect morphological classifications from citizen science projects can increase the efficiency in which new stripping candidates are found. Projects such as Galaxy Zoo can aid in the identification of ram pressure stripped galaxies that are key to understanding galaxy evolution in clusters.}


   \maketitle
%

\section{Introduction}
\label{sec:intro}

There is increasing evidence that ram pressure stripping (hereafter RPS) is one of the main drivers of galaxy transformations in dense environments such as galaxy clusters \citep[e.g.,][]{Boselli:2006a, Cortese:2021a, Boselli:2022a} and maybe even lower density groups \citep[e.g.,][]{Rasmussen:2006a, Vulcani:2018a, Roberts:2021b, Vulcani:2021a, Kolcu:2022a, Roberts:2022a}. As a galaxy enters a group or cluster, the gas within the galaxy experiences a pressure due to the intracluster medium. This pressure, first described in \citet{Gunn:1972a}, can heat up and strip the gas within infalling galaxies, removing the gas from the galaxies and preventing star formation. The hydrodynamic removal of the gas from the infalling galaxy can occur within a cluster crossing time \citep[e.g.,][]{Jaffe:2015a, Jaffe:2018a, Smith:2022a} and even as fast as a few hundred Myr \citep[e.g.,][]{Vollmer:2004a, Bekki:2009a, Boselli:2016a}, which eventually leads to a full quenching of star formation in dense environments \citep[e.g.,][]{Vulcani:2020a, Cortese:2021a, Boselli:2022a, Vulcani:2022a}.

There are many ways to observe a galaxy that is experiencing ram pressure stripping. Trails of gas and dust are stripped from the galaxy, with visible deformation seen in the H\scaleto{$I$}{1.2ex} gas \citep[e.g.,][]{Gavazzi:1984a, Gavazzi:1995a, Crowl:2005a, Chung:2007a, Chung:2009a}. In some cases, this stripped gas can condense and form stars behind the galaxy as it moves through the cluster, which can be seen in UV and optical imaging \citep[e.g.,][]{Smith:2010a, Abramson:2011a, Merluzzi:2013a, Poggianti:2016a, Fossati:2016a}. This star formation, as well as shock heating, excite the surrounding gas in the wake of the galaxy, producing bright H$\alpha$ tail-like features \citep[e.g.,][]{Gavazzi:2001a, Sun:2007a, Yagi:2007a, Yagi:2010a, Merluzzi:2013a, Poggianti:2017a, Gavazzi:2017a}.

Gas can also be compressed along the leading edge of the galaxy, leading to visible star formation on one side of the galaxy \citep[e.g.,][]{Vogt:2004a,Rasmussen:2006a, Poggianti:2019a, Ramatsoku:2019a, Ramatsoku:2020a, Vulcani:2022a, Roberts:2022b}. Cosmic rays from the resultant supernovae can also be stripped from the galaxy, giving asymmetric radio continuum morphologies analogous to the optical tails \citep[e.g.,][]{Roberts:2021a, Roberts:2021b, Roberts:2022a}. The most spectacular and dramatic galaxies experiencing RPS are often known as `jellyfish' galaxies for the long trails of gas and star formation (e.g.,\ \citealt{Chung:2009a, Smith:2010a}, see also \citealt{Bekki:2009a}), with features that can often be seen across multiple wavelengths \citep[e.g.,][]{Crowl:2005a, Sun:2007a, Crowl:2008a, Merluzzi:2013a, Poggianti:2019a}. 

Detailed studies of these galaxies have uncovered a wealth of information about the transformations that occur while a galaxy experiences RPS. For example, ram pressure stripped galaxies are likely to have a short-term burst in star formation before quenching \citep[e.g.,][]{Bekki:2003a, Kapferer:2009a, Vulcani:2018b, Roberts:2020a, Cortese:2021a}, and a higher than average efficiency of converting H\scaleto{$I$}{1.2ex} into H$_{\textrm{2}}$ (e.g.,\ \citealt{Moretti:2020a}; \citealt{Villanueva:2022a},  see also \citealt{Brown:2023a}; \citealt{Watts:2023a}). Stripped galaxies also contain highly ordered magnetic fields \citep[][]{Muller:2021a}, and potentially have a high likelihood of hosting an active central black hole \citep[e.g.,][]{Poggianti:2017b, George:2019a, Ricarte:2020a, Peluso:2022a}. However, these effects are not seen across all ram pressure stripped galaxies (e.g., \citealt{Boselli:2022a}, see also discussions from \citealt{Cattorini:2023a}).

While analysis of many of these ram pressure stripped galaxies has greatly improved our knowledge of this process, we are still limited to studies of relatively small samples of the order of tens of  galaxies \citep[e.g.,][]{Smith:2010a, Yagi:2010a, McPartland:2016a, Ebeling:2019a, Roberts:2020a, Gullieuszik:2020a}. There are few examples of studies which have studied hundreds of galaxies experiencing RPS \citep[e.g.,][]{Poggianti:2016a, Roberts:2021a, Durret:2021a, Roman-Oliveira:2021a, Boselli:2022a}, with only $\sim1000$ RPS candidate galaxies currently known across all studies in the published literature. While we have learned a great deal about many individual galaxies, only a large, statistical sample of ram pressure stripped galaxies will enable us to understand the properties of ram pressure stripped galaxies as a population. 

The limitation to creating such samples of stripped galaxies is the difficulty in identifying such a population. The features of ram pressure stripped galaxies are often very faint, and the tails, gas compressions, and warping of gas can have a variety of shapes across different wavelengths in galaxy imaging. Several studies have used a combination of CAS parameters \citep{Conselice:2003a}, Gini and M$_{20}$ values \citep{Lotz:2004a}, shape asymmetry \citep{Pawlik:2016a}, star formation asymmetry, and centroid offsets to identify RPS candidates \citep{McPartland:2016a, Roberts:2020a, Troncoso-Iribarren:2020a, Roberts:2021a, Roberts:2021b, Bellhouse:2022a, Liu:2021a, Krabbe:2024a}. However, visual identification of RPS features is still one of the primary and most reliable methods of identifying such galaxies. 

There has been some success with neural networks in classifying morphological features in galaxies, including disks, spirals arms, and bars \citep[e.g.,][]{Dieleman:2015a, Huertas-Company:2015a, Dominguez-Sanchez:2018a, Walmsley:2020a, Bekki:2021a, Walmsley:2022a}. However, the low numbers of known ram pressure stripped galaxies means potential training sets could suffer from over-fitting \citep[e.g.,][]{Dieleman:2015a, Huertas-Company:2015a}. Additionally the diversity of RPS features that have been observed means neural networks may not yet be the best tool for finding such rare and diverse objects \citep[e.g., see][for similar discussions on merging galaxies]{Lambrides:2021a}. It is problems such as these where citizen science may hold the answer to finding these rare, visually distinctive objects.

One well known example of a morphological citizen science project is the Galaxy Zoo project \citep[e.g.,][]{Lintott:2008a, Willett:2013a}. The Galaxy Zoo project and subsequent follow up programs have facilitated the study of visual morphological features of galaxies for hundreds of thousands of objects. This includes investigations into the arms of spiral galaxies \citep[e.g.,][]{Hart:2017a, Hart:2018a}, the presence and lengths of bars \citep[e.g.,][]{Melvin:2014a, Simmons:2014a}, and probing the dynamics of mergers \citep[e.g.,][]{Darg:2010a, Darg:2010b, Holincheck:2016a}. These projects have also found rare extra-galactic objects that would likely be missed in automated classifications. These include Green Peas \citep[e.g.,][]{Cardamone:2009a}, the ionised clouds near AGN \citep[such as Hanny's Voorverp,][]{Keel:2012a}, and other rare objects found through Galaxy Zoo talk forums which have since been further investigated with the Hubble Space Telescope \citep[e.g.,][]{Keel:2022a}.

While the relationship between environment and galaxy morphology has been explored with Galaxy Zoo \citep[e.g.,][]{Bamford:2009a, Darg:2010b, Skibba:2012a, Smethurst:2017a}, and ram pressure stripping galaxies have been identified using citizen science for simulated galaxies \citep[][]{Zinger:2024a}, there has so far been very little investigation into identifying observed galaxies experiencing RPS. Ideally, citizen scientists could be tasked with identifying features such as tails, compressions, and spiral deformations which are consistent with RPS. While such studies are forthcoming, we aim to investigate whether existing Galaxy Zoo classification can separate galaxies experiencing RPS from other galaxies.

This study analyses the morphological classifications of known ram pressure stripped galaxies, to determine whether general citizen science programs can be used to better select ram pressure stripped galaxies. Several previous studies have identified likely RPS candidates in clusters that are within the footprint of the Galaxy Zoo projects \citep[e.g.,][]{Smith:2010a, Yagi:2010a, Poggianti:2016a, Roberts:2020a, Roberts:2021a}. If these galaxies have systematic morphological differences to other galaxies, it may be true that new candidates can be found by selecting galaxies with such morphological parameters. We can then use expert classifications to test whether such Galaxy Zoo morphology motivated samples contain new RPS candidates, and analyse them in more detail.

The outline of this manuscript is as follows: In Sect.\ \ref{sec:sample} we describe the sample properties, including sourcing known ram pressure stripped galaxies. In Sect.\ \ref{sec:GZmorphologies} we compare the morphological classifications of these ram pressure stripped galaxies to comparison samples of galaxies. We then use these morphological classifications to motivate a new potential sample of ram pressure stripped galaxies in Sect.\ \ref{sec:GZ_motivated}. We discuss our results in Sect.\ \ref{sec:discussion}, before concluding in Sect.\ \ref{sec:conclusion}. Throughout this manuscript, unless otherwise stated, we use AB magnitudes, $1\sigma$ binomial confidence interval uncertainties, and assume a concordance cosmology for all luminosity distances, absolute magnitudes, and cluster R$_{200}$ values, with $\Omega_{m} = 0.3$, $\Omega_{\Lambda} = 0.7$, and H$_{0} = 70$ km s$^{-1}$ Mpc$^{-1}$.

\section{Sample}
\label{sec:sample}

\subsection{Galaxy Zoo 2}
\label{sec:GZ2}
In order to investigate the ability of citizen science classifications in characterising RPS features, we need a large homogeneous citizen science data set from which to draw our sample. The Galaxy Zoo project \citep[][]{Lintott:2008a} was a large project tasked with classifying the morphologies of over 200,000 galaxies in data release 6 of the Sloan Digital Sky Survey \citep[SDSS,][]{Adelman-McCarthy:2008a}. The initial project used a very simple classification scheme, intended on separating galaxies into spirals and ellipticals. This was extended in Galaxy Zoo 2 \citep[][]{Willett:2013a}, which involved a more detailed classifications of a subset of the SDSS Data Release 7 \citep[][]{Abazajian:2009a}. This sample has a brighter magnitude limit, but involves a much more extensive morphological classification, including questions asking about the presence of a bar, whether the galaxies have a certain number of spiral arms, and if there are any odd features in the galaxy \citep[see][for the full question description]{Willett:2013a}. This last question is the most critical to this study, as ram pressure stripped galaxies tend to have very visually distinct morphologies. 

We take the Galaxy Zoo 2 classifications \citep[hereafter GZ2,][]{Willett:2013a}, and use the debiasing of \citet{Hart:2016a}, which contains the most up-to-date methodology for debiasing voter fractions for changes in redshift. This is done to account for the lower resolution of high redshift galaxies, where less detailed features can be seen. This method debiases vote fractions using analytic functions based on a low redshift sample of Galaxy Zoo classified galaxies, and can more effectively debias classifications in multiple response questions compared to previous methods \citep[see][for details]{Hart:2016a}. This dataset contains morphological analysis of $\sim 240,000$ galaxies within the SDSS survey above a magnitude limit of $r = 17.0$, up to a redshift of $z = 0.25$. These were matched with redshift and photometric information from SDSS DR7 \citep[][]{Abazajian:2009a} for each galaxy for use in this study.

\subsection{Local ram pressure stripped galaxy sample}
\label{sec:samp}
In this study we are investigating the morphologies of known ram pressure stripped galaxies to determine whether they are different to other galaxy populations. Therefore, we require a large sample of known ram pressure stripped galaxies from the literature to form our ram pressure stripped galaxy sample. We draw our sample from over $\sim 30$ studies that visually identify RPS features in cluster and group galaxies. These studies all identify features such as tails of debris and star formation, and unilateral asymmetries to identify galaxies experiencing RPS. The full list of studies from where our sample is drawn is shown in Table \ref{tab:known_cand_list}. All of these studies list galaxies that have been visually identified to contain features consistent with ram pressure stripping, and thus form the basis of this sample.

These galaxies have been found in a variety of ways, including broadband optical data, UV imaging, H$\alpha$ emission, H\scaleto{$I$}{1.2ex} gas, and 144MHz radio continuum. The combined catalogue contains a sample of $>900$ ram pressure stripped galaxies, which was used to match with morphological data, and will be published in a follow-up study. We note that this catalogue contains any visually identified galaxy with signatures consistent with RPS. This includes galaxies where the cluster membership may not be known, as well as galaxies with small visual disturbances and hard to determine origins. Some of the galaxies used in this study may not be true ram pressure stripped galaxies, but have an appearance consistent with RPS. We therefore note that these galaxies are ram pressure stripping candidates (or RPS candidates), and for many sources, further validation is required to unambiguously confirm that ram pressure stripping is the dominant process.

\begin{table*}
	\centering
	\caption{Studies from which ram pressure stripped candidates are sourced in this study.}
	\label{tab:GZ_jf_sources}
	\begin{tabular}{|l|c|c|c|} 
		\hline
		Study & Cluster & Detection method & No. of RPS candidates with GZ2 data \\
		& & (instrument) & (total including doubles) \\
		\hline
		\citet{Gavazzi:1984a} & Abell 1367 & 1.4GHz (VLA) & 1 \\
		\citet{Conselice:1998a} & Coma & Optical (WIYN) & 1 \\
		\citet{Yagi:2007a} & Coma & H$\alpha$ (Suprime-Cam) & 1 \\
		\citet{Smith:2010a} & Coma & UV (Galex) & 6 (10) \\
		\citet{Yagi:2010a} & Coma & H$\alpha$ (Suprime-Cam) & 10 (13) \\
		\citet{Kenney:2015a} & Coma & Optical (Hubble) & 1 \\
		\citet{Poggianti:2016a}  & Wings Clusters & Optical (WINGS) & 37 \\
		\citet{Poggianti:2016a}  & PM2GC Groups & Optical (MGC) & 24 \\
		\citet{Gavazzi:2017a} & Abell 1367 & H$\alpha$ (Suprime-Cam) & 1 \\
		\citet{Roberts:2020a}  & Coma & Optical (CFHT) & 20 (41) \\
		\citet{Roberts:2021a}  & Various clusters & 144MHz (LOFAR) & 63 (66) \\
		\citet{Roberts:2021b} & Various groups & 144MHz (LOFAR) & 47  \\
		\hline
	\end{tabular}
    \label{tab:known_cand_list}
    \tablefoot{
 The values in the final column denote how many galaxies have corresponding Galaxy Zoo 2 data. The values in brackets indicate the number of galaxies including duplicates, where a galaxy is present in multiple studies.
}
\end{table*}

A subset of this sample contains morphological data from GZ2. This includes many of the galaxies seen in \citet{Poggianti:2016a}, as well as those in studies from \citet{Roberts:2020a, Roberts:2021a, Roberts:2021b}. Many of these galaxies have been identified as having ram pressure stripping features with deeper imaging than is available to those in Galaxy Zoo, and as such may not have been classified in Galaxy Zoo. The full list of references are shown in Table \ref{tab:GZ_jf_sources}.

A total of 212 galaxies were found to match a GZ2 counterpart, and are considered in this sample. While many more galaxies fall within the SDSS footprint and redshift range, several galaxies (especially those in the Virgo cluster) do not contain GZ2 morphological classifications. The matched sample includes galaxies with H$\alpha$ tails, visible B and R band trails of star formation, filaments of CO and dust emission, as well as asymmetries in 144MHz radio continuum emission (see Table \ref{tab:GZ_jf_sources} for details). We use all of these galaxies for analysis in the following sections.

In $\sim100$ cases in our sample, no assessment of RPS in the optical bands has been conducted. In such cases, two of the co-authors (JC and YJ) inspected the optical morphology using $grz$ colour images from the DESI Legacy Imaging Surveys \citep[hereafter Legacy Survey,][]{Dey:2019a}. This included giving a ranking on a scale of 1-5 of the strength of the stripping features, assessing how obvious any potentially stripped features appear, and how confident a classifier is of the presence of RPS \citep[as done in ][]{Poggianti:2016a}. While using broad band optical imaging does not always give a full picture of potential RPS, it does allow a reasonable assessment of optical deformations in ram pressure stripped galaxies (e.g., \citealt{Poggianti:2016a}, \citealt{Durret:2021a}, \citealt{Vulcani:2022a}, see also \citealt{Kolcu:2022a}). 

The sample of known RPS candidates, along with associated cluster and redshift information, GZ2  morphologies, and photometric data from SDSS DR7 form our primary sample for use in Sect. \ref{sec:GZmorphologies}. These galaxies span a redshift range of $0.011< z < 0.12$ with a median of $z = 0.035$, and an approximate absolute magnitude range of $-22.82 < M_{r} < -16.81$ with a median of $M_{r} = -20.38$.

\section{Morphologies in Galaxy Zoo}
\label{sec:GZmorphologies}
\subsection{Creating matched comparison samples}
\label{sec:comparisons}
In this study we aim to compare the morphologies of known ram pressure stripped galaxies to other (non-stripped) galaxies in GZ2. However, this comparison is likely affected by many biases - GZ2 galaxies are based on a magnitude limited SDSS sample, while many of the ram pressure stripped candidates are often large, high surface brightness galaxies, where detailed features of RPS can be seen. We need to ensure we do not compare large, bright stripped galaxies to faint and small cluster or field galaxies.

Therefore, we create samples of comparison galaxies to mitigate differences in redshift and magnitude between the ram pressure stripped galaxies and the general GZ2 sample. We create two separate comparison samples, one consisting of similar magnitude galaxies within clusters, and another consisting of similar magnitude galaxies in the field. Each galaxy in these two comparison samples is matched in both absolute $r$ band magnitude and redshift to a single galaxy in the known RPS sample. We describe the selection of these samples below.

The cluster comparison sample is comprised of Galaxy Zoo galaxies that lie close to known cluster centres from \citet{Tempel:2014a}. We take the centroid positions of all clusters and calculate on-sky and velocity differences to Galaxy Zoo galaxies. We define a GZ2 galaxy as being a cluster member if it lies within 4R$_{200}$ and $4\sigma$ of the cluster velocity dispersion from any cluster centre (similar to the cluster membership criteria of the known RPS sample). Galaxies that are associated with more than one cluster centre are assigned to the nearest centre on the sky. In this process we only consider galaxies that are within clusters of mass  $>10^{14}M_{\odot}$, and thus do not consider small groups of galaxies.

Galaxies must only be within the 4R$_{200}$ and $4\sigma$ radius and velocity limits - strict membership within the catalogue of \citet{Tempel:2014a} is not required. This approach aims to best select galaxies that lie within cluster cores as well as infall regions, where features of RPS may be expected. Selecting galaxies from only the cluster cores would likely miss many infalling galaxies similar to the known RPS galaxies, which could potentially bias our results \citep[e.g.,][]{Vollmer:2001a, Bellhouse:2017a, Rhee:2017a, Jaffe:2018a}.

As a second comparison, we create a sample of field galaxies that are not associated with any group or cluster in the catalogue of \citet{Tempel:2014a}. A GZ2 galaxy is defined as a field galaxy if it is not within 5R$_{200}$ and $5\sigma$ velocity offset from a group or cluster centre with mass $>10^{12}M_{\odot}$ in \citet{Tempel:2014a}. In this way galaxies considered to be field are not likely associated with a cluster or group. Galaxies in this field sample will act as a baseline, to determine whether any morphological properties are common in cluster galaxies, or specific to those experiencing RPS.

To make our comparison samples, we find eight galaxies that match the absolute $r$ magnitude and redshift of each known RPS galaxy, within a small $\Delta M_{r}$ and $\Delta z$. In cases where more than eight galaxies are found within these limits, we select the eight galaxies that have the smallest $\Delta M_{r}$ to their respective known RPS galaxy. Each cluster comparison galaxy matches the corresponding known RPS galaxy within $|\Delta M_{r}| < 0.45$ and $|\Delta z| < 0.1$. The field comparison galaxies have magnitudes and redshifts that are within $|\Delta M_{r}| < 0.06$ and $|\Delta z| < 0.01$ of each respective known RPS galaxy. The different $\Delta M_{r}$ and $\Delta z$ values between the two comparison samples is due to a smaller number of galaxies within large clusters compared with field galaxies. We note that as there is a high number of known RPS candidates found in local clusters, a large $\Delta z$ has been used. Both comparison samples contain a total of 1696 galaxies ($212 \times 8$, see Table \ref{tab:samp_list} for an overview of these samples).

\begin{figure}
	\resizebox{\hsize}{!}{\includegraphics{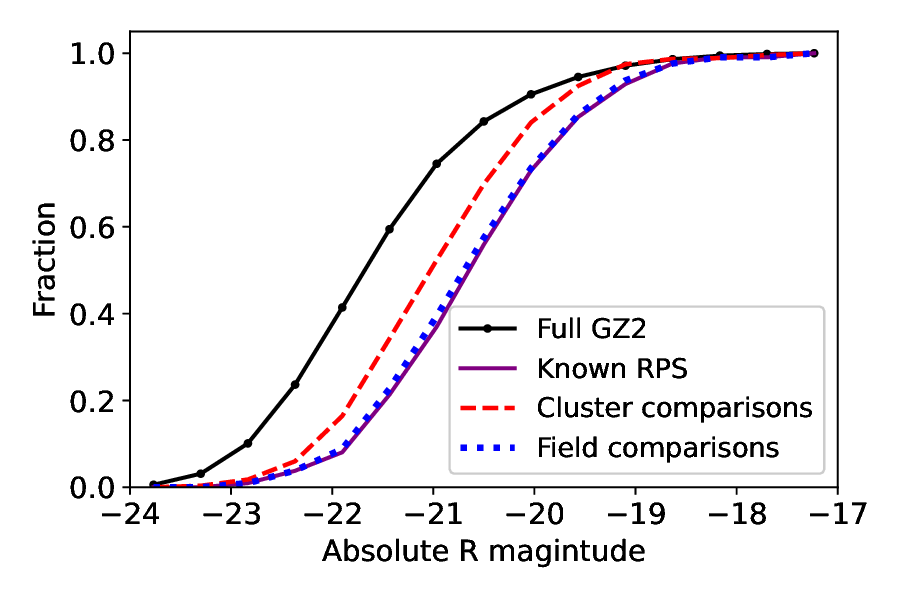}}
	\resizebox{\hsize}{!}{\includegraphics{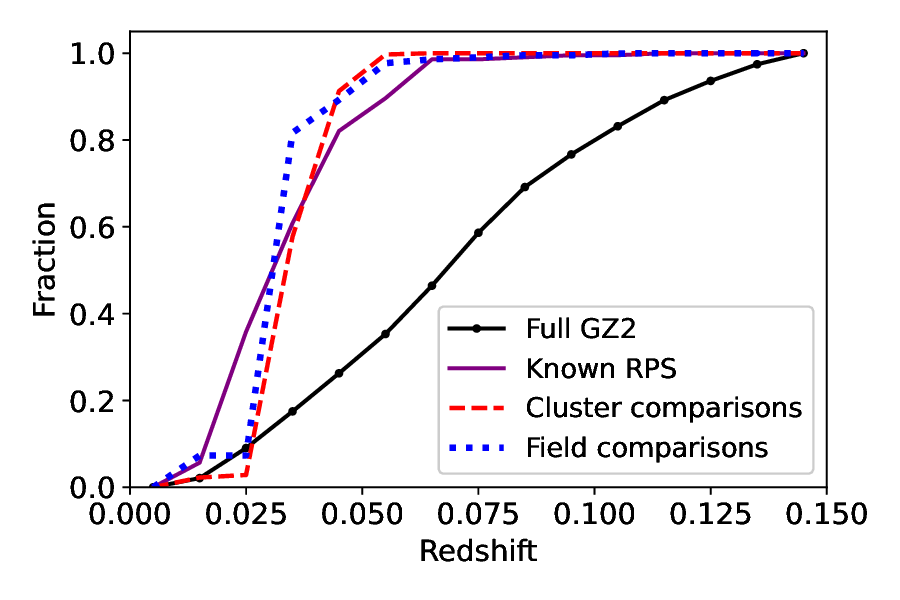}}
    \caption{Cumulative distributions of the absolute $r$ band magnitude (top) and redshift (bottom) for the known RPS Sample (magenta), field comparison sample (blue), and cluster comparison sample (red). The full Galaxy Zoo dataset is shown in black. The distributions are largely similar between the comparison samples and the known RPS sample, however, there is a lack of low redshift objects to match with Coma Cluster RPS candidates.}
    \label{fig:compsample_stats}
\end{figure}

Cumulative distributions of the absolute $r$ magnitude and redshift for the known RPS, cluster comparison, and field comparison samples are shown in Fig.\ \ref{fig:compsample_stats}. The full Galaxy Zoo sample is included for reference. While best efforts have been made to match the three samples in both redshift and absolute magnitude, there is a significant difference between the known RPS galaxy sample and both of the comparison samples in the distribution of redshifts. This is due to the known RPS sample containing many galaxies within the Coma Cluster at a redshift of $z=0.0029$. As there are few galaxies within this redshift range, the known RPS sample has a sharp rise at very low redshift, which cannot be perfectly replicated in the comparison samples. However, while there is a difference in the distribution, the analysed redshift range is small enough that there is no significant difference in the surface brightness in these samples, suggesting the three samples do not significantly differ in their properties due to redshift evolution.

We note that no effort is made to match the colour of the RPS candidate sample to the comparison samples. While the comparison samples have similar absolute $r$ band magnitude distributions, the comparison samples include many red galaxies, which are generally absent in the RPS candidate sample. We discuss the effect of controlling for colour in Sect.\ \ref{sec:col-matched}. Additionally, our principal findings are unchanged if we consider a similar cluster comparison sample of galaxies within 1.5R$_{200}$ and $2\sigma$ of a cluster core, instead of 4R$_{200}$ and $4\sigma$.

\subsection{Morphologies of known RPS galaxies}
\label{sec:morphologies}
With suitable control samples defined for each of our known RPS candidates, we now compare the morphologies of the known RPS candidates with the two comparisons samples, based on the vote distributions in GZ2. 

\begin{figure*}
\sidecaption
	\includegraphics[width=12cm]{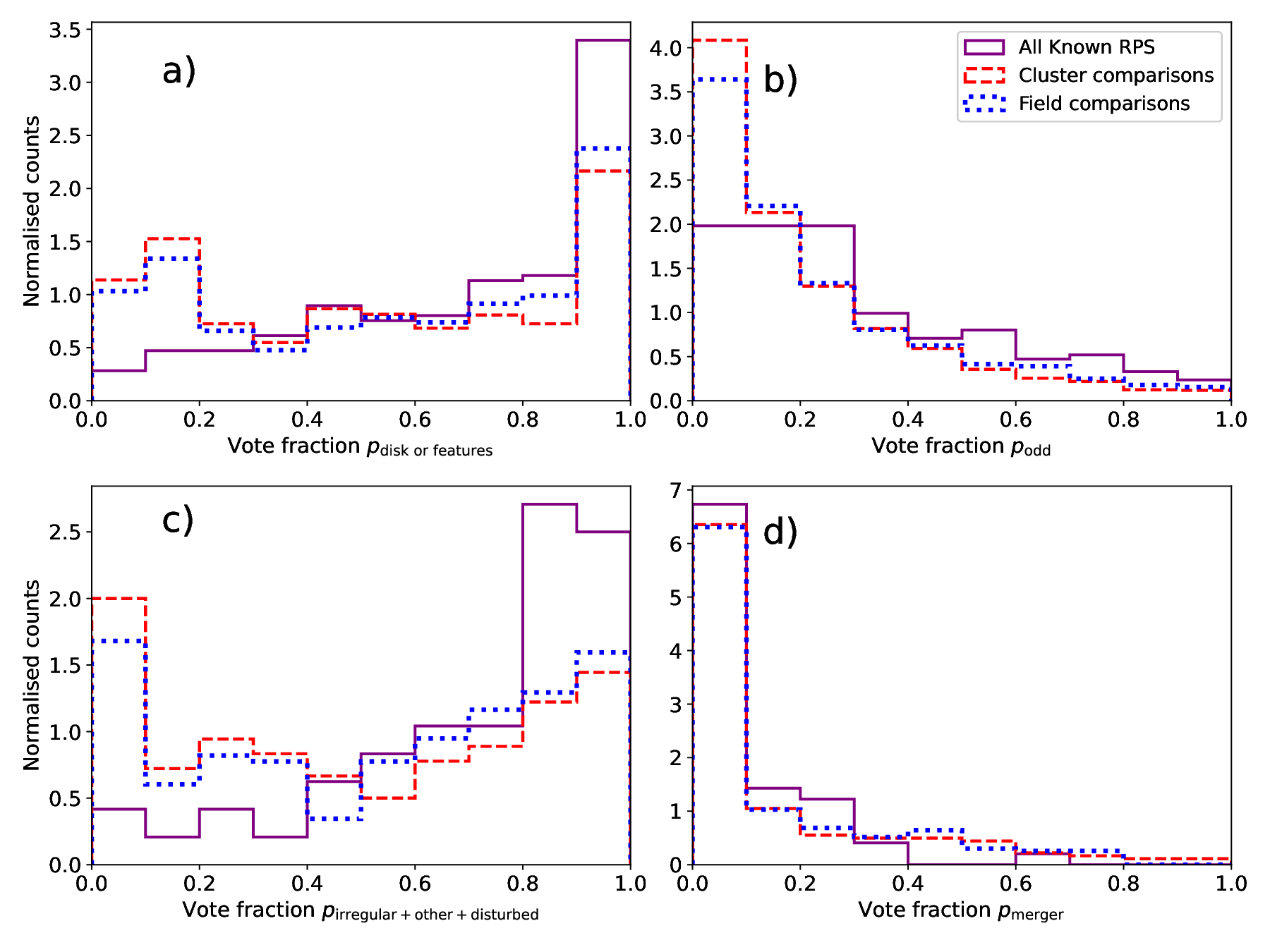}
	    \caption{
	    a) Debiased fraction of votes that consider a galaxy to have features or a disk. Compared to the cluster and field comparison samples, the known RPS candidates have a much higher proportion of disk galaxies, in line with expectations.
	    b) Debiased fraction of votes for a galaxy to be considered odd. Compared to the group comparison sample, the known RPS candidates have a much higher proportion of odd candidates.
	    c) debiased fraction of votes for galaxies to have `irregular', `disturbed', or `other' features, for galaxies with $p_{\textrm{odd}} > 50\%$. 
	    d) debiased fraction of votes for galaxies to have `merger' features, for galaxies with $p_{\textrm{odd}} > 50\%$. Compared to either the group or field comparison samples, the ram pressure stripped galaxies are considered to be disturbed or irregular, and are not considered to be mergers. All counts are normalised such that the area under each curve is equal to one.}
    \label{fig:GZ_multi_all}
\end{figure*}

We first investigate the overall morphology of the samples, based on responses to the question `Is the galaxy simply smooth and rounded, with no sign of features or a disk?'. We show in panel a) of Fig.\ \ref{fig:GZ_multi_all} the fraction of votes for each galaxy to exhibit `features or disk' morphologies in GZ2. These fractions have been `debiased' to account for differences in redshift \citep[see][for details]{Hart:2016a}. A higher debiased vote fraction indicates a higher proportion of people thought that a given galaxy had features or a disk (denoted as $p_{\textrm{disk}}$). We see that galaxies in the known RPS sample have a higher fraction of disk-dominated and featured galaxies, compared with the two comparison samples. In the known RPS sample $72 \pm 3\%$ of galaxies have $p_{\textrm{disk}}> 0.5$ compared with the field and cluster comparison samples of $57 \pm 1\%$ and $51 \pm 1\%$ respectively. Additionally, the vote distribution of the known RPS galaxies is significantly different to either comparison sample, with Kolmogorov--Smirnov test (K-S test), finding a significant difference with a p-value $< 1\times10^{-10}$. This is not unexpected, as the known RPS sample is selected from galaxies with trails of gas and star formation present, which are more common in spiral galaxies \citep[e.g., results from][]{Poggianti:2016a, Roberts:2020a, Kolcu:2022a}. We also see that both the cluster and field comparisons have similar distributions of galaxies with `features or disk' morphologies. This is potentially due to the cluster comparison sample containing galaxies in cluster infall regions, with some tidal and lenticular galaxies considered to have `features or disk', despite potentially being a traditional `early-type' galaxy.

In panel b) Fig.\ \ref{fig:GZ_multi_all} we compare the distribution of votes for each galaxy in response to the question `Is there anything odd?'. We see that the known RPS galaxy sample has a higher fraction of galaxies with high odd vote fractions compared with the comparison samples. While the field and cluster comparisons have $14 \pm 1 \%$ and $11 \pm 1 \%$ galaxies with $p_{\textrm{odd}} > 50\%$, the known RPS sample has over $23 \pm 3\%$. The distributions are also significantly different, with the K-S test finding a p-value $< 1\times10^{-8}$. This shows that the trails of ram pressure induced star formation are potentially being noticed as an odd feature by galaxy zoo volunteers, despite the shallow SDSS imaging. 

We show the breakdown of odd feature classifications for all galaxies with $p_{\textrm{odd}} > 50\%$ in panel c) and d) of Fig.\ \ref{fig:GZ_multi_all}. Voters can select from several different `odd' feature possibilities, including `ring', `disturbed', `irregular', `merger', and `other' \citep[see][for the full list]{Willett:2013a}. A high fraction of known RPS candidates are considered `disturbed', `irregular', and `other', with $82 \pm 6\%$ of the known RPS candidates with $p_{\textrm{odd}} > 50\%$ having $p_{\textrm{irregular + disturbed + other}} > 50\%$ (see panel c of Fig.\ \ref{fig:GZ_multi_all}). This contrasts the field and cluster comparison samples, which have significantly smaller irregular+disturbed+other fractions for galaxies with $p_{\textrm{odd}} > 50\%$ ($58 \pm 3\%$ and $49 \pm 4\%$ respectively). Additionally, we see in panel d) of Fig.\ \ref{fig:GZ_multi_all} that no galaxy in the known RPS sample is considered to be undergoing a `merger', highlighting the ability of galaxy zoo classifiers to distinguish these ram pressure features as different from merger features.

Previous studies have observed that galaxies experiencing RPS can have a possible `unwinding' effect on galaxies, with an apparent teasing out of spiral arms as they infall \citep[e.g.,][]{Bellhouse:2021a, Vulcani:2022a}. While this is not confirmed to be a ram pressure stripping driven phenomenon, we can test whether the known RPS sample appears to have a population of unwinding arm galaxies. GZ2 contains information to quantify the `tightness' of the spiral arms in the question `How tightly wound do the spiral arms appear?', with responses being either `tight', `medium', or `loose'. This can be used as a proxy to test the possible unwinding of spiral arms seen in galaxies experiencing RPS. \citet{Masters:2019a} defines an arm winding score, based on the debiased vote fractions to this question. The arm winding score is defined as:
\begin{equation}
w_{avg} = 0.5p_{\textrm{arms medium}} + 1.0p_{\textrm{arms tight}}    
\label{equ:armwind}
\end{equation}
where $p_{\textrm{arms medium}}$ and $p_{\textrm{arms tight}}$ are the debiased vote fractions for the responses `medium' and `tight' respectively. For this analysis, we only consider galaxies in each of the three samples which have $p_{\textrm{disk}} > 0.43$, $p_{\textrm{has spiral arms}} > 0.619$, and $p_{\textrm{not edge-on}} > 0.715$, as per \citet{Willett:2013a} and \citet{Masters:2019a}.

\begin{figure}
	\resizebox{\hsize}{!}{\includegraphics{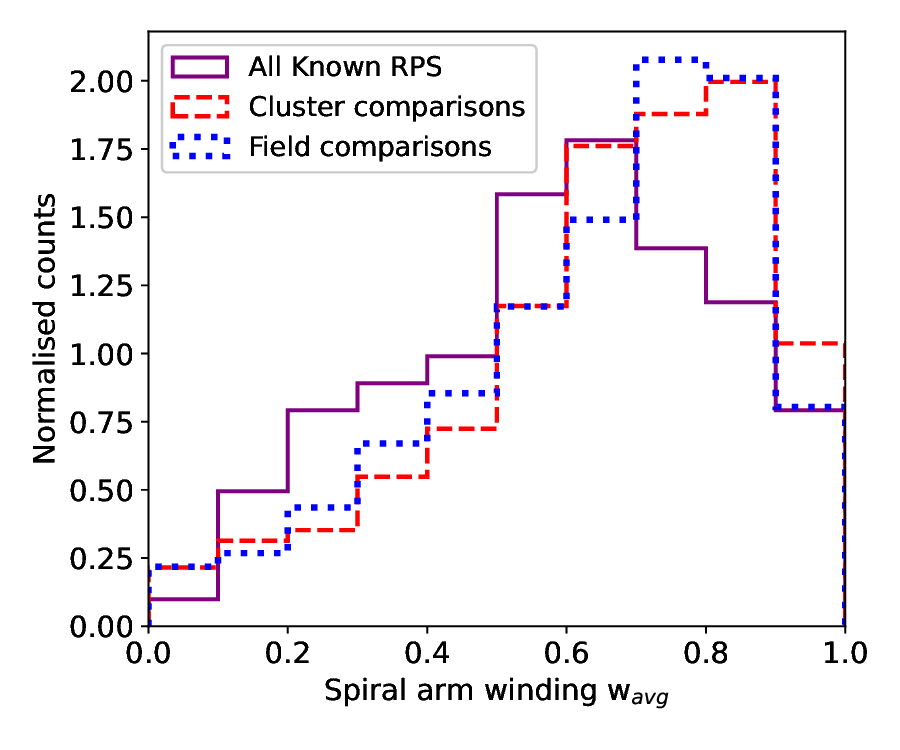}}
    \caption{Spiral arm winding, w$_{avg}$, for all face-on spiral galaxies, calculated from Equation \ref{equ:armwind}. Compared to the comparison samples, the known RPS candidates have lower average arm winding, indicating a more loose spiral arm pattern.}
    \label{fig:GZ_arm_hist_all}
\end{figure}

In Fig.\ \ref{fig:GZ_arm_hist_all} we show the breakdown of the spiral arm winding, w$_{avg}$, for all non edge-on galaxies that are considered spiral. The known RPS candidates have a significantly lower average arm winding, compared with either the cluster or field comparison samples, with a K-S test p-value of $\sim0.005$. This lower winding score is indicative of a more loosely wound spiral arm pattern. It may be that this is a manifestation of the unwinding phenomenon described in \citet{Bellhouse:2021a} and \citet{Vulcani:2022a}. We discuss this result, and unwinding spiral arms in clusters in more detail in Sect.\ \ref{sec:unwind}.

A recent study by \citet[][]{Sanchez-Garcia:2023a} finds a link between the central star formation in ram pressure stripped galaxies with the existence of a galaxy bar, and find that the highest enhancement in star formation occurs predominantly in barred ram pressure stripped galaxies. It may be that the star formation enhancements seen in ram pressure stripping might be more easily observed in barred galaxies. We compare the fraction of known ram pressure stripped galaxies that contain a bar based on the response to the question 'Is there a sign of a bar feature through the centre of the galaxy?'. For this we only consider face-on spiral galaxies, with $p_{\textrm{disk}} > 0.43$, and $p_{\textrm{not edge-on}} > 0.715$ as per \citet{Willett:2013a}. 

We find there is no significant increase in the fraction of barred galaxies between the known RPS sample and the two comparison samples. The fraction of known RPS galaxies with a strong bar, defined as $p_{\textrm{bar}} > 0.5$ \citep[e.g.,][]{Skibba:2012a}, is $23 \pm 4\%$. This value is very similar to the field comparisons with a bar fraction ($25 \pm 2\%$) and on average lower than the cluster comparison sample ($32 \pm 2\%$), with a K--S test having a difference of p-value of $0.03$.

These values are largely consistent with \citet{Skibba:2012a}, who found a bar fraction of $25.3 \pm 0.4\%$ in their Galaxy Zoo sample. If we consider a more generous criterion of $p_{\textrm{bar}} > 0.2$ we find bar fractions of $53 \pm 4\%$, $52 \pm 2\%$ and $62 \pm 2\%$ for the known RPS, field comparisons, and cluster comparisons respectively. The results show a lower bar fraction in the known RPS galaxies compared to the cluster comparison sample, but a similar fraction to other field galaxies. Therefore, we do not find evidence that this sample of RPS galaxies is more likely to host a bar than our magnitude matched cluster and field comparison galaxies, and in fact may be lower than that of similar magnitude cluster galaxies.

\subsection{Strong ram pressure stripping candidates}
\label{sec:strong_morphologies}
The sample used in the previous section considers all possible cases of RPS. This includes those with only mild stripping features as well as galaxies identified with CO disk deformations and asymmetries in the 144MHz radio continuum emission. While many of these galaxies may also have optical signatures of RPS, there is not necessarily a one-to-one overlap (see Sect. \ref{sec:radio_only} for details). Therefore, we rerun the previous results considering only the RPS cases which are known to have optical RPS features. Where appropriate, we also remove galaxies that exhibit low RPS strength.

Defining the `strength' of RPS is an inherently subjective task, so we caution that what constitutes `strong' RPS will differ from study to study, and classifier to classifier. We note that the sample used in this study does not necessarily constitute a definitively pure `strong' RPS sample. However, it is still useful to compare the results from the previous section to galaxies that have the most prominent and obvious RPS features.

We then restrict our sample to only include galaxies that have previously been identified with optical features. In cases where a strength of RPS was given, we only include galaxies that have a higher strength of RPS \citep[typically $\ge3$ on a 5 point scale, e.g.,][]{Poggianti:2016a}. In studies where the optical/UV RPS strength was not measured, the strength classifications described in Sect.\ \ref{sec:samp} were used, and only the galaxies with a strength value $\ge3$ were considered. This leaves a sample of 58 known RPS candidates. We therefore only consider the corresponding comparison galaxies in each of the comparison samples, such that there are 464 cluster and 464 field comparison galaxies.

\begin{figure}
    \resizebox{\hsize}{!}{\includegraphics{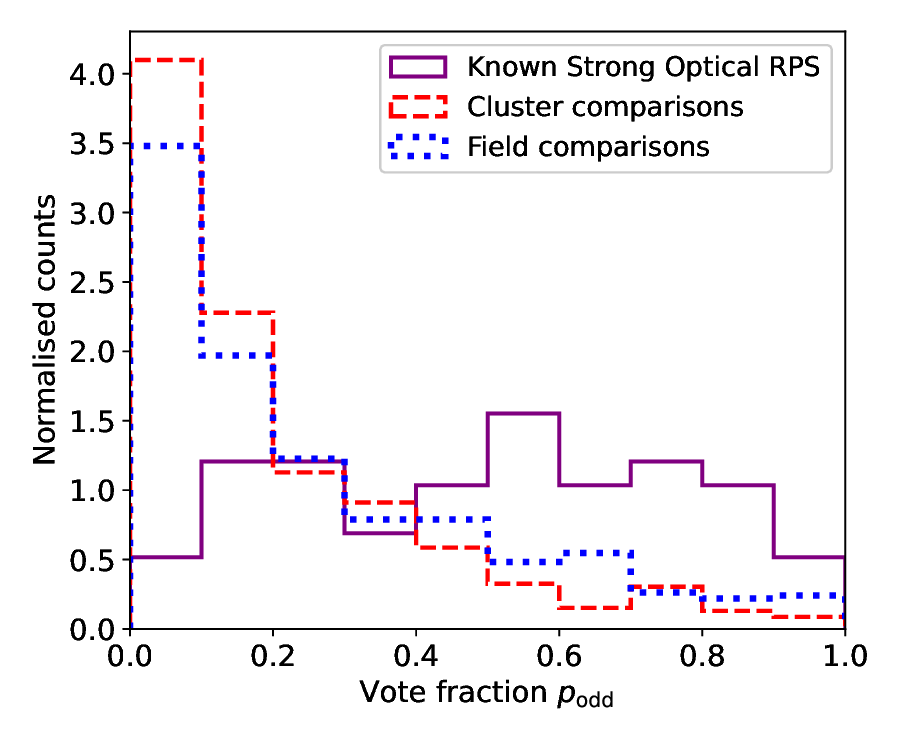}}
	\resizebox{\hsize}{!}{\includegraphics{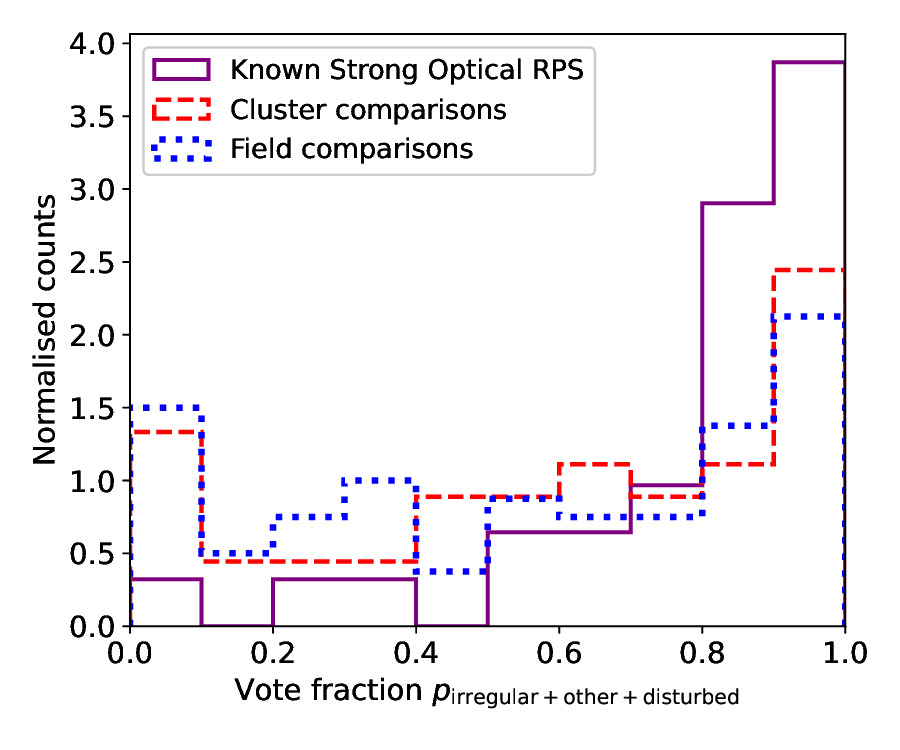}}
    \caption{Top: Debiased vote fraction for strong ram pressure galaxies displaying odd features. The strong known RPS galaxies have a higher fraction of `odd' features compared to the comparison samples. 
    Bottom: Debiased vote fraction for galaxies having `irregular', `disturbed', or `other' features, considering only strong ram pressure stripped galaxies with $p_{\textrm{odd}} > 50\%$. These galaxies are more likely to have these morphological classifications compared to the matched comparison samples. The counts are normalised such that the area under each curve is equal to one.}
    \label{fig:GZ_odd_breakdown_highJ}
\end{figure}

We find qualitatively similar results to Sect.\ \ref{sec:morphologies}  when we consider only strong optical RPS galaxies. The strong optical RPS galaxies tend to be more disk-dominated, are likely to have odd features, and have loose spiral arm winding when compared with either cluster or field comparisons. However, we notice that many of the trends are much stronger than those seen in Sect. \ref{sec:morphologies}. For example, we find $53 \pm 7\%$ of the strong known RPS galaxies have $p_{\textrm{odd}} > 50\%$, compared with $17 \pm 2\%$ and $10 \pm 1\%$ for their respective field and group comparisons. The distribution is seen in Fig.\ \ref{fig:GZ_odd_breakdown_highJ}, and is different from either comparison sample with a p-value $< 1\times10^{-10}$. The difference between the samples represents a factor of $\sim2.3$ increase in the number of objects with $p_{\textrm{odd}} > 50\%$ compared to the full known RPS sample.

Additionally, the fraction of known RPS candidates described as being `disturbed' or `irregular' also increases. The lower panel of Fig.\ \ref{fig:GZ_odd_breakdown_highJ} shows the fraction of objects with $p_{\textrm{odd}} > 50\%$ which are considered to be either `disturbed', `irregular', or `other'. We find that $90 \pm 5\%$ of the odd, strong known RPS galaxies have $(p_{\textrm{disturbed}} + p_{\textrm{irregular}} + p_{\textrm{other}}) > 50\%$, compared with $59 \pm 6\%$ and $65 \pm 7\%$ in the field and cluster comparison samples respectively. The K-S test shows the distributions are different with a p-value $=0.002$. While the trends are similar to those seen in the previous section, we find that there is a much higher fraction of odd candidates when only considering candidates that display strong ram pressure features, suggesting that citizen scientists can better notice ram pressure stripping features in expert defined `strong' ram pressure stripping galaxies.

\subsection{Morphologies of radio-only RPS galaxies}
\label{sec:radio_only}
The previous results show that on average our sample of ram pressure stripped galaxies have different morphologies compared with non-stripped field and cluster comparisons. This includes many galaxies with H$\alpha$ tails and/or optically visible debris, as well as galaxies with H\scaleto{$I$}{1.2ex}, CO or radio continuum asymmetries. However, investigating only the latter class of RPS galaxy might yield different results to the full sample. 

For example, \citet{Roberts:2021a, Roberts:2021b} find asymmetries in the 144MHz radio continuum of LOFAR galaxies to identify RPS candidates. These galaxies are not selected on their optical morphology, and may not share the same morphological traits as other ram pressure stripped galaxies. As gravitational perturbations are unlikely to produce a radio continuum asymmetry without a corresponding optical deformation, these  galaxies may represent a more `pure' sample of ram pressure stripped objects. These galaxies may also be in a different stage of stripping compared to galaxies that also have an optical deformation. Here we investigate the optical morphologies of the known ram pressure stripped galaxies that have been identified using non-optical data.

We select 134 known ram pressure stripped galaxies from our original 212, taking only those which are detected at 144MHz, and compare these with their respective field and cluster comparison galaxies. As in Sect.\ \ref{sec:morphologies}, we find an excess disk fraction in the radio-identified RPS galaxies compared with the comparison samples. However, unlike the previous results, we see a smaller enhancement in the odd fractions of radio ram pressure stripped galaxies. The radio-identified RPS candidates have only $21 \pm 4\%$ of objects with $p_{\textrm{odd}} > 50\%$. While still larger than the cluster and field comparison samples ($13 \pm 1\% $and $10 \pm 1\%$ respectively), the difference is much smaller. Additionally, we do not see a large enhancement in the `disturbed' or `other' types of odd galaxies compared with the field and cluster comparisons.

We note that many of these galaxies with radio RPS signatures also have RPS features that are visible in optical imaging. For example, many of the galaxies with ram pressure stripped features in the 144MHz band from \citet{Roberts:2021a} also have RPS features in other bands including UV, H$\alpha$, and optical \citep[from][respectively]{Smith:2010a, Yagi:2010a, Roberts:2020a}. Many other radio selected RPS galaxies had optical RPS features identified through the classifications described in Sect. \ref{sec:samp}. 

However, there remains a significant fraction of radio-identified ram pressure stripped galaxies without any optical signatures of RPS. In our sample there are 75 galaxies with no optical signatures reported either in other studies, or by our internal classification. This represents $35\%$ of our sample, indicating that many galaxies are likely experiencing RPS without displaying obvious optical RPS features (See Table \ref{tab:samp_list} for a comparison of samples).

We repeat our GZ2 classifications analysis only considering the 75 galaxies without optical RPS features. We find that the number of galaxies where $p_{\textrm{odd}} > 50\%$ is now lower in the radio RPS sample than the comparisons. These radio only RPS galaxies have a $p_{\textrm{odd}} > 50\%$ fraction of $6 \pm 3\%$, which is lower than both of the field and cluster comparisons ($11 \pm 2\%$ and $11 \pm 1\%$ respectively). This difference has a small significance, with a K-S test p-value of 0.06. This would suggest that galaxies without optical RPS features may look odd less frequently than other field or cluster galaxies.

This result is unsurprising, given each galaxy was individually inspected to not have any ram pressure features in optical imaging. However, as the expert classification only considered RPS features, it could be possible that the galaxy has a different odd feature (from a tidal interaction, merger, etc.). Given the fraction of odd objects greatly drops when optical RPS features are excluded, it is potential indirect evidence that the GZ2 classifiers are primarily noticing the RPS features when denoting an object as `odd'. This result also shows that galaxies without obvious optical trails of star formation can still be undergoing RPS, and a single wavelength study is unlikely to find the full population of ram pressure stripped galaxies.

\section{Citizen science to identify ram pressure stripping}
\label{sec:GZ_motivated}
\subsection{Galaxy Zoo motivated sample}
\label{sec:motivated}
In the previous section, we found that known ram pressure stripped galaxies have morphologies that are distinct from other cluster and field galaxies. These traits, while not unique to RPS galaxies, are distinct when compared to galaxies of a similar magnitude and redshift. Therefore, it may be possible to use these morphological characteristics to identify new samples of disturbed galaxies, including potential new ram pressure stripped galaxy candidates. Selecting galaxies that have a combination of these morphological features may help narrow down the search for new RPS candidates, compared with simply visually inspecting all galaxies around clusters.

To test this hypothesis, we create a sample of galaxies from GZ2, with GZ defined morphological features consistent with the known RPS galaxies. We select Galaxy Zoo classified galaxies with at least 20 classifications, that are within 1.5R$_{200}$ of any cluster with mass $>10^{14}M_{\odot}$ from \citet{Tempel:2014a}. We adopt a smaller radius than in Sect. \ref{sec:GZmorphologies} to better search in cluster centres, where ram pressure stripped galaxies are commonly found \citep[e.g.,][]{Jaffe:2018a}. Additionally, we require that they have a cluster velocity offset of $v/\sigma < 4$. We also consider only galaxies with a blue optical colour, satisfying $(g - i) < 1.05$, which corresponds to the reddest $(g - i)$ colour in the known RPS sample. 

With these blue cluster galaxies selected, we then restrict this sample to have morphological parameters that match the known RPS galaxies. We select galaxies with the following debiased vote fractions:

\begin{itemize}
    \item $p_{\textrm{disk}} > 0.5$ AND
    \item $p_{\textrm{odd feature merger}} < 0.1$ AND
    \item $p_{\textrm{not edge-on}} > 0.5$ AND
    \item $p_{\textrm{arms loose}} > 0.2$ OR
    \item $p_{\textrm{arms loose}} + $ $p_{\textrm{arms medium}} > 0.72$
\end{itemize}
we also considered galaxies with:
\begin{itemize}
    \item $p_{\textrm{disk}} > 0.5$ AND
    \item $p_{\textrm{odd}} > 0.5$ AND 
    \item $p_{\textrm{odd merger}} < 0.1$ AND
    \item $p_{\textrm{odd irregular}} + $ $p_{\textrm{odd disturbed}} > 0.5$
\end{itemize}

These values were chosen to reflect the results of Sect. \ref{sec:morphologies}, selecting galaxies with the strongest odd disturbed and irregular features, as well as loose wound spiral arms, whilst ensuring a sample size appropriate for visual expert morphological classifications. Selecting galaxies using these morphological features yields 319 morphologically selected galaxies out of 1237 blue disk dominated cluster galaxies. We call this new sample `GZ-selected' galaxy sample, to denote this sample of cluster galaxies with morphological classifications described above, as our primary sample for further analysis.

\subsection{Non-GZ selected control sample}
\label{sec:controls}
In order to make a comparison with this new GZ-selected galaxy sample, we create a control sample of cluster galaxies. These control galaxies have the same constrains in cluster membership, and (g - i) colour as the GZ-selected sample in Sect. \ref{sec:motivated}. Additionally, all of these controls have $p_{\textrm{disk}} > 0.5$. However, unlike the GZ-selected sample, none of the controls fulfil the criteria listed in Sect. \ref{sec:motivated}. In this way, we can compare whether using specific morphological criteria can boost the fraction of RPS galaxies found using expert classifications.

In order to match the GZ-selected sample, we select 319 galaxies that are matched within $|\Delta M_{r}| < 0.35$ and $|\Delta z| < 0.12$; one control for each of the GZ-selected candidates (see Table \ref{tab:samp_list} for reference). In this way, we have even numbers of GZ-selected and non-GZ-selected cluster galaxies. This will allow us to determine whether specific Galaxy Zoo based morphological classifications can more efficiently identify new ram pressure stripped galaxies than other spiral galaxies.

\subsection{GZ-field sample}
\label{sec:false}
Ram pressure stripping requires a dense intergalactic medium, and as such only affects galaxies in dense environments such as galaxy clusters and groups. Any morphological deformations in galaxies outside of these dense environments are unlikely due to ram pressure stripping. Therefore, creating a sample of galaxies away from clusters can give an estimate of the contribution of `false positive' ram pressure stripped galaxies detected via this method. 

We create a sample of field GZ2 galaxies as an extra comparison sample. We select Galaxy Zoo galaxies that are $>5R_{200}$ and $5\sigma$ velocity offset from a group or cluster centre of any mass in \citet[][analogous to Sect. \ref{sec:comparisons}]{Tempel:2014a}. From this we only consider galaxies that have GZ2 classifications that match those in Sect.\ \ref{sec:motivated}. This gives a sample of $\sim9700$ field galaxies, with GZ2 classification consistent with the results of Sect.\ \ref{sec:GZmorphologies} to create a field sample.

We then create a matched sample of 319 galaxies, analogous to the controls in Sect. \ref{sec:controls}. These galaxies are all matched to respective GZ-selected galaxies within $|\Delta M_{r}| < 0.02$ and $|\Delta z| < 0.05$. We call this sample the GZ-field sample.

In addition to these three samples of cluster galaxies, we also select 100 galaxies from the known RPS sample with optical ram pressure stripping features to use as an upper limit. The four samples: the GZ-selected cluster sample, the non-GZ control cluster sample, the GZ-field sample, and the known RPS sample are all taken for expert visual classification. To avoid confusion, we use the phrase `expert classifiers' and `expert classifications' to distinguish these from Galaxy Zoo classifications.

\subsection{Visual expert classification of candidate samples}
With these samples created, expert classifiers individually analyse each galaxy in the GZ selected sample, the non-GZ-selected controls, the known optical RPS galaxies, and the GZ-field sample using colour images from Legacy Survey \citep[][]{Dey:2019a}. The layout for this expert classification is described in Table \ref{tab:GASP_questions}. We note that these Legacy Survey images are deeper than the SDSS images of the original GZ2 dataset. There will therefore be some features visible that were not able to be identified by the GZ classifiers. However, we assume that almost all features identified in the GZ imaging will be visible in the Legacy Survey images.

This questionnaire was uploaded into the Zooniverse project creator\footnote{\url{https://www.zooniverse.org/lab}} to facilitate classifications. This project was then sent to expert classifiers, who are all familiar with RPS morphologies, features, and identification. The galaxies were assessed based on whether they appeared to have visible signatures of ram pressure stripping. The primary discriminator for a galaxy to be considered ram pressure stripped is the first question listed in Table \ref{tab:GASP_questions}. This question directly asks the expert classifier if the galaxy has a morphology consistent with a galaxy experiencing ram pressure stripping. The features that the expert classifiers identified include one-sided tails/debris trails, compression of gas, or knots of potential star formation along one side of the galaxy \citep[e.g.,][see Table. \ref{tab:GASP_questions} for details]{McPartland:2016a,Poggianti:2016a,Roberts:2020a,Roman-Oliveira:2021a,Durret:2021a,Kolcu:2022a, Roberts:2022a}. Additionally, the expert classifiers were also asked to consider whether there is an unwinding of the spiral arms, similar to the search conducted by \citet{Vulcani:2022a}. 

If a galaxy was considered to be experiencing ram pressure stripping, it was then further classified according to the observed features. The strength of the ram pressure stripping was also measured, analogous to the $J_{\textrm{class}}$ seen in \citet{Poggianti:2016a}. If a tail/direction of motion was able to be estimated, the classifier was also asked to draw the estimated direction of motion with a line.

\begin{figure}
	\resizebox{\hsize}{!}{\includegraphics{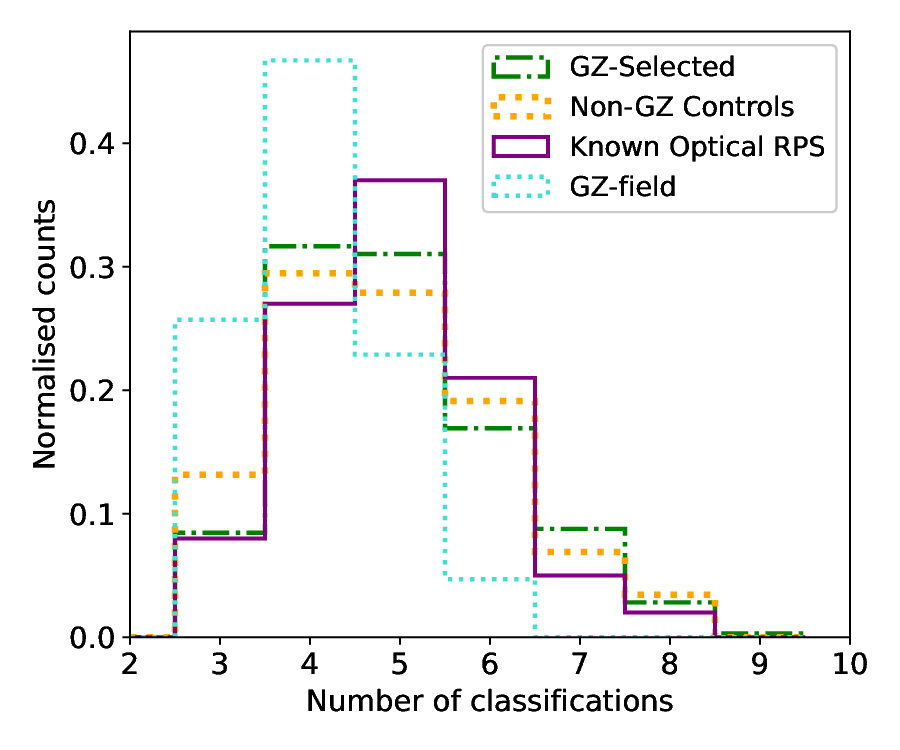}}
    \caption{Number of votes a galaxy received during the expert classification. Each galaxy was observed a minimum of three times, with an median of five experts classifying any individual galaxy.}
    \label{fig:GASP_classification_3samp}
\end{figure}

Each galaxy was then classified by several experts, in order to determine whether these galaxies are potential ram pressure stripping candidates. Each galaxy was classified a minimum of three times, with a median of five classifications and a maximum of ten classifications. The histogram of the number of expert classification per galaxy is shown in Fig. \ref{fig:GASP_classification_3samp}. 

\begin{figure}
	\resizebox{\hsize}{!}{\includegraphics{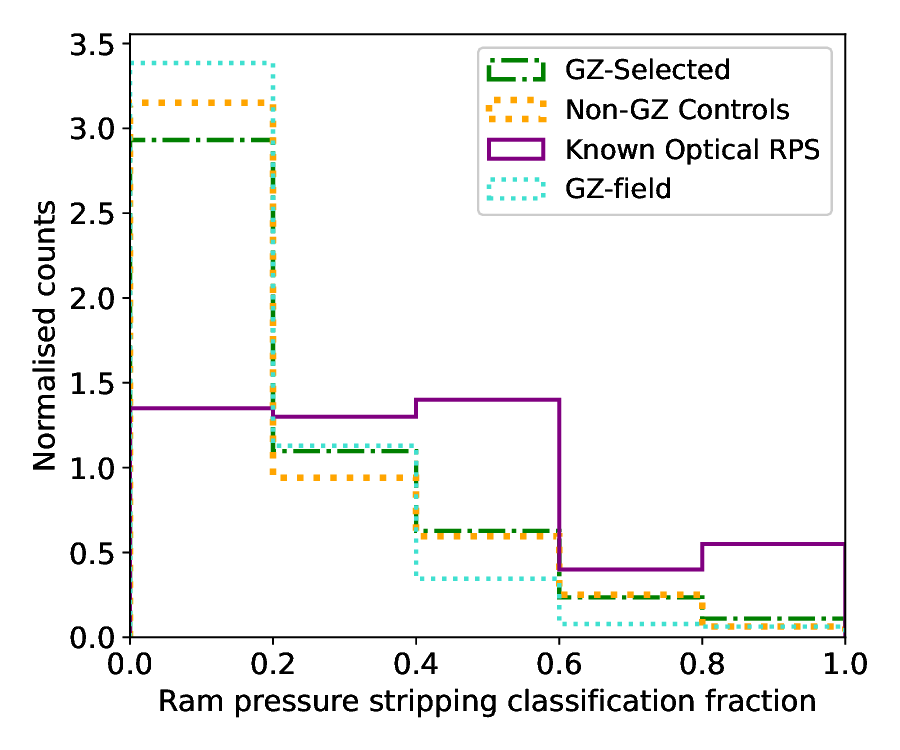}}
	\resizebox{\hsize}{!}{\includegraphics{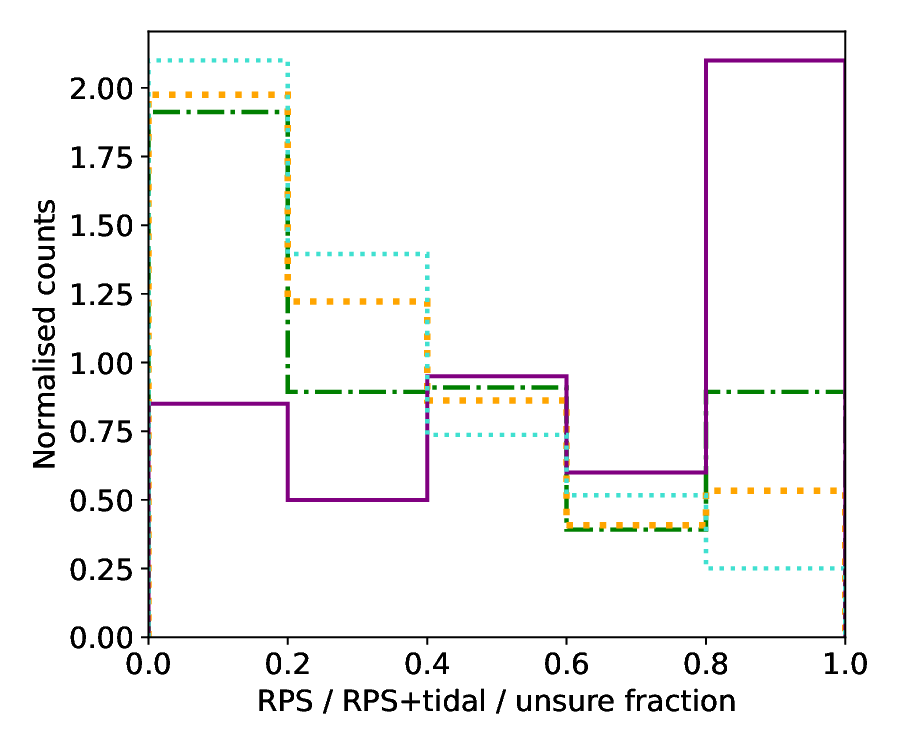}}
    \caption{Top: Fraction of votes considering a galaxy to be affected by RPS. Based on expert classifications, a low number of both the GZ-selected and non-GZ controls have clear RPS features. Bottom: Fraction of votes for RPS, both RPS and tidal features, or features of undetermined origin. The known RPS galaxies have a higher fraction of objects that are considered to have RPS features when considering the three categories, compared to only RPS (see top panel). Using this metric, the GZ-selected sample (green) contains more galaxies with a high fraction of RPS votes, compared with the non-GZ controls (orange), or the GZ-field (cyan). All counts are normalised such that the area under each curve is equal to one.}
    \label{fig:GASP_rps_votes}
\end{figure}

In Fig.\ \ref{fig:GASP_rps_votes} we compare the fraction of expert classifier votes that considered a galaxy to be experiencing ram pressure stripping. The initial question asks whether there is an unusual feature present in the galaxy. The top panel shows the distribution of votes for the response `It looks like the galaxy has tails caused from ram pressure stripping'. We see that the GZ-selected sample and the non-GZ-selected sample both have similar distributions, with only 33 and 26 objects in each respective samples having more than $50\%$ of expert votes suggesting a given galaxy is experiencing ram pressure stripping (hereafter denoted f$_{\textrm{RPS}}> 0.5$ to separate the expert vote fraction compared with Sect.\ \ref{sec:morphologies}). This is not a significant difference, with a confidence interval difference $<1\sigma$, and a K--S test not finding a difference to $1\sigma$. 

We see that very few of the GZ-field sample have f$_{\textrm{RPS}}> 0.5$, with only 11 objects considered as such. This is not unexpected, as these galaxies are not in dense cluster regions, and thus are not expected to be experiencing RPS. Interestingly, there are also very few of the known RPS galaxies that have a vote fraction above f$_{\textrm{RPS}}> 0.5$ ($27\%$). The low f$_{\textrm{RPS}}$ is likely due to a combination of factors. Firstly, there are limitations of using a single three colour cutout image of the galaxies when conducting an expert classification. In many of the past studies, deeper imaging was used \citep[e.g.,][]{Kenney:2015a, Roberts:2020a} or a more detailed inspection of each galaxy was conducted \citep[e.g.,][]{Poggianti:2016a}. Faint features seen in a deep narrow band fits image may not be seen in a 3-colour cutout. A second factor is the inclusion of a combination of tidal/RPS, and for cases where the origin is ambiguous. Some classifiers may prefer these options when dealing with cases of RPS that are not completely obvious.

In the lower panel of Fig.\ \ref{fig:GASP_rps_votes} we show the distribution of votes considering three responses that could indicate RPS. In this figure we combine the fraction of votes for RPS, RPS with a tidal interaction simultaneously, and a feature that the classifier is  unsure (denoted f$_{\textrm{RPS/tidal/unsure}}$). When we use f$_{\textrm{RPS/tidal/unsure}}$, we see a small increase in the fraction of galaxies with 97 objects with f$_{\textrm{RPS/tidal/unsure}}> 50\%$ in the GZ-selected sample compared with 76 in the non-GZ-selected controls. This is also true of the GZ-field sample, with 55 galaxies with f$_{\textrm{RPS/tidal/unsure}}> 50\%$. The K--S test suggests this difference is around $1\sigma$, so while stronger than the pure RPS fraction, the difference in this distribution alone does not show a significant difference.

Additionally, 63 out of 100 known RPS galaxies have high f$_{\textrm{RPS/tidal/unsure}}$, suggesting that many of the expert classifiers may have been unsure of the origin of a morphological deformation. This highlights that even for known RPS galaxies, there can be cases of ambiguity as to whether the morphological features in optical images are ram pressure stripping. 

\subsection{Expert classified new candidates}
\label{sec:new_cands}
In the previous section, experts classified the morphologies of different samples of galaxies, in order to identify potential ram pressure stripping features. Here we define the metric for new ram pressure stripping candidates using the expert classifications, in order to test whether the use of GZ2 classifications in preselection can more efficiently find possible RPS galaxies. We also test the false positive rate of finding RPS features using the field sample.

Using the tallied expert classifications, we create a sample of new RPS candidates from the subsamples. We adopt a set of criteria based on f$_{\textrm{RPS}}$ and f$_{\textrm{RPS/tidal/unsure}}$ for the voted galaxies in the GZ-selected, non-GZ control, and GZ-field samples. As seen in Fig. \ref{fig:GASP_rps_votes}, the known RPS galaxies were often found to have an ambiguous deformation, or combination of RPS as well as a tidal interaction. We therefore create our sample considering both of these possibilities. This sample of ``expert classified RPS candidates'' are defined as galaxies with at least three expert votes, and the fraction of expert votes such that:
\begin{itemize}
    \item f$_{\textrm{RPS}} > 0.5$ OR
    \item f$_{\textrm{RPS/tidal/unsure}} > 0.75$
\end{itemize}

This classification scheme prioritises galaxies where more than half of the expert classifiers thought the galaxy has visual signatures of ram pressure stripping. Additionally, it also considers galaxies to have ram pressure stripping in cases where over three quarters of experts have classified galaxies as either ram pressure stripping, a combination of ram pressure stripping and tidal interactions, or the classifier is unsure of the origin of any morphological deformation. Our criteria include both of these fractions as many of the known RPS galaxies have a low f$_{\textrm{RPS}}$, but a high f$_{\textrm{RPS/tidal/unsure}}$ (see Fig. \ref{fig:GASP_rps_votes}). We discuss changes to these values in Sect. \ref{sec:Selections}. We note that while these galaxies are expert classified, they are purely `candidates', and further investigation of a combination of the ionised H$\alpha$ emission, the H\scaleto{$I$}{1.2ex} gas, local orbital dynamics, and/or stellar population modelling of the tails would be required to confirm if the morphologies of these galaxies are predominantly due to ram pressure stripping.

\begin{figure*}
    \begin{centering}
	\includegraphics[width=17cm]{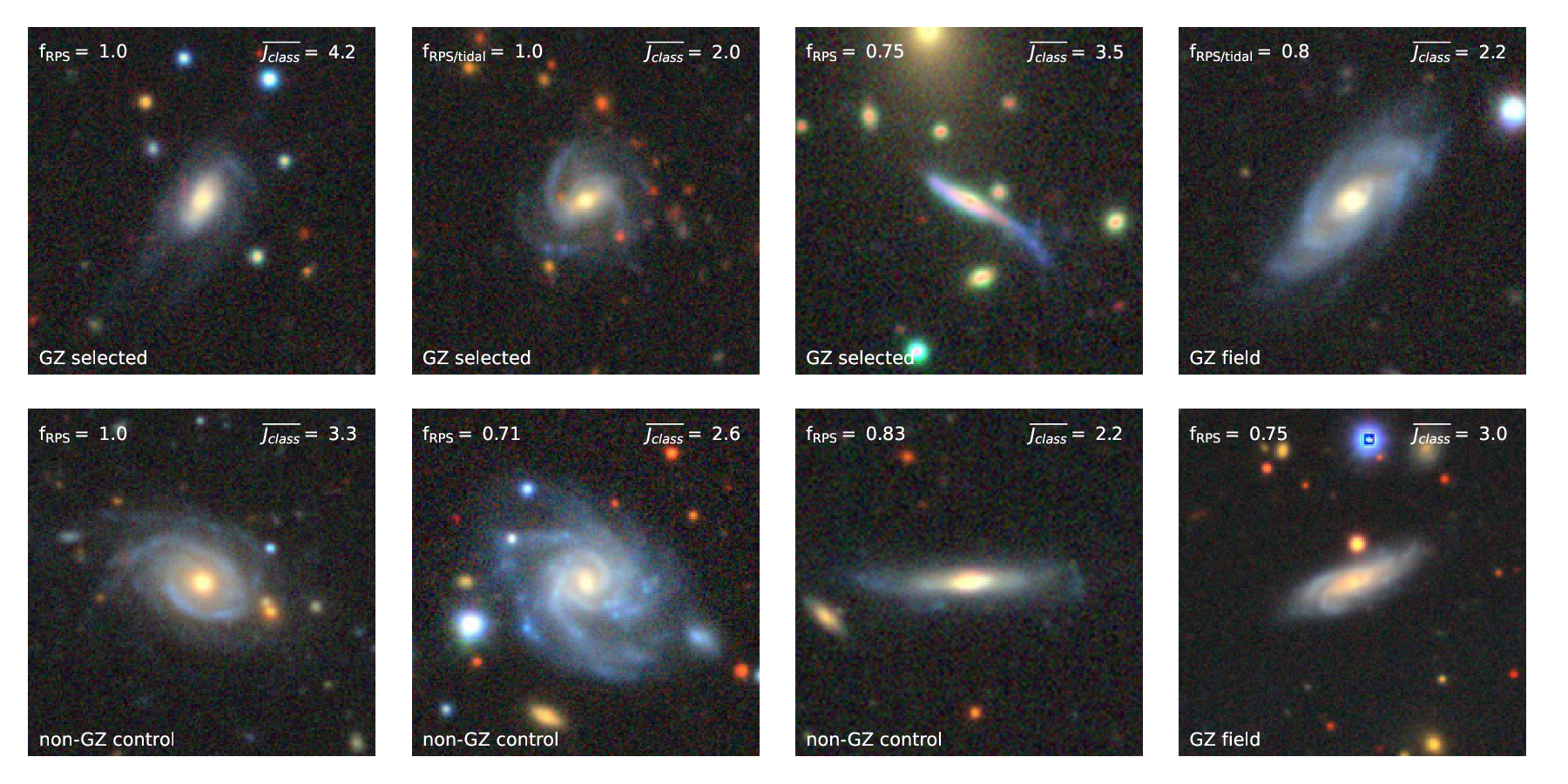}
    \caption{Legacy Survey 3 colour images of several expert classified RPS candidates found in this study, including galaxies from the GZ-selected sample, the non-GZ control sample, and the GZ-field candidates. The f$_{\textrm{RPS}}$ value is shown if f$_{\textrm{RPS}} > 0.5$, and the f$_{\textrm{RPS/tidal/unsure}}$ is shown if f$_{\textrm{RPS}} \le 0.5$. The average $J_{\textrm{class}}$ value, $\overline{J_{class}}$, value is also shown. Further examples are also included in Fig. \ref{fig:extra_cand_images}}
    \label{fig:new_cand_images}
    \end{centering}
\end{figure*}

Figure \ref{fig:new_cand_images} shows examples of the new candidates found through this method. The images are example candidates from the GZ-selected sample, the non-GZ controls, and the GZ-field sample. These galaxies all have different features, with tails, unwinding spiral arm patterns, and star forming regions that are consistent with ram pressure stripping. Additional examples are also included in Fig. \ref{fig:extra_cand_images}.

In addition to classifying whether a galaxy was experiencing ram pressure stripping, the expert classifiers also characterised any potential RPS seen, and what types of RPS features were present. The classifiers rated the strength of any potential RPS on a scale of 1-5 (denoted $J_{\textrm{class}}$, see \citealt{Poggianti:2016a} for details), with one being indicative of a minor disturbance, and five being typical of the most striking and impressive jellyfish galaxies in the literature \citep[e.g.,][see also Sect. \ref{sec:samp}]{Poggianti:2016a}. We define a sub-sample of `strong' RPS candidates as candidates which satisfy the previous criteria, but also have $\overline{J_{class}} \ge 3$, where $\overline{J_{\textrm{class}}}$ is the mean $J_{\textrm{class}}$ value taken from all the classifiers. Examples of both strong candidates with $\overline{J_{class}} \ge 3$, and candidates with $\overline{J_{class}} < 3$ are displayed in Fig. \ref{fig:new_cand_images} with several extra examples in Fig. \ref{fig:extra_cand_images}. 

Using the definitions of ram pressure stripping previously described, we find a total of 101 expert classified RPS candidates (see Table \ref{tab:samp_list}). This includes 62 candidates from the GZ-selected sample, as well as 39 candidates in the control sample. 24 of these are considered `strong' candidates with $\overline{J_{class}} \ge 3$, and 77 candidates that have $\overline{J_{class}} < 3$.

Additionally, we find 19 galaxies in the GZ-field sample that meet the same criteria as the expert classified RPS candidates. This also includes 10 galaxies with $\overline{J_{class}} \ge 3$. Examples of these galaxies are included in Fig. \ref{fig:new_cand_images} and Fig. \ref{fig:extra_cand_images}. 

In the subsequent analysis, we compare these GZ-field galaxies to both the GZ-selected and non-GZ control samples to place a limit on likely contamination from other morphological features. However, as these galaxies are not considered close to a known SDSS group/cluster, we do not include this sample in the total number of new RPS candidates, or include these in Table \ref{tab:samp_list}).

\subsubsection{Ram pressure stripping fraction}
\label{sec:GZ_frac}
\begin{figure}
	\resizebox{\hsize}{!}{\includegraphics{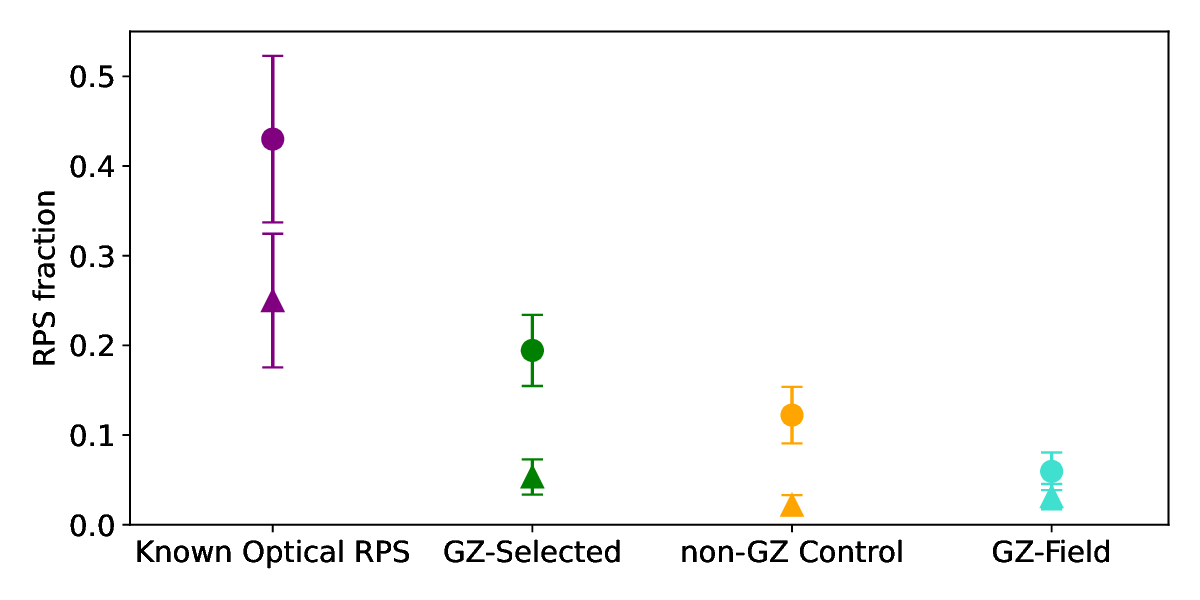}}
    \caption{Fraction of galaxies considered to be expert classified RPS candidates in each of the three samples defined in Sect.\ \ref{sec:motivated}. The fraction of expert classified RPS is shown with a circle, while the fraction of new strong candidates, with $\overline{J_{class}} \ge 3$, is shown with a triangle. Binomial one sigma confidence intervals are also shown. We see that known ram pressure stripped candidates have the highest fraction. Additionally, the GZ motivated sample contains a higher fraction than the control sample.}
    \label{fig:GASP_rps_frac}
\end{figure}

With the expert classified RPS candidate sample created, we can now assess whether Galaxy Zoo morphological classifications can more efficiently find galaxies that are likely to be experiencing ram pressure stripping. We show in Fig. \ref{fig:GASP_rps_frac} the fraction of expert classified RPS candidates (circles) and new strong RPS candidates (triangles) for the four samples created in Sect.\ \ref{sec:motivated}. We see that, as expected, the known RPS sample has a significant fraction of galaxies that are considered to be strong RPS candidates by this metric. However, we note that not all of the known RPS candidates are classified as RPS galaxies by our metric. This is due to differences in the imaging used in this study. While the 3-colour Legacy Surveys imaging is deep, does not always allow the faint deformations to be seen as found in WINGS or the Canada-France-Hawaii telescope \citep[][]{Poggianti:2016a,Roberts:2020a}.

When we compare the RPS fraction of the GZ-selected and the non-GZ control samples, we find that the GZ-selected sample has a significantly higher fraction of expert classified RPS candidates. As seen in Fig.\ \ref{fig:GASP_rps_frac}, $19 \pm 2\%$ of the GZ-selected sample are expert classified RPS candidates, 1.5 times higher than the non-GZ control candidates ($12\pm 2\%$). This represents a $>2\sigma$ confidence interval increase in the fraction of RPS candidates in the GZ sample, when compared to the control sample. This shows that using Galaxy Zoo morphological features can be a helpful approach to more efficiently find RPS candidates. We note that while the GZ-selected sample has been designed to have morphologies consistent with known RPS galaxies, the non-GZ control sample still has an RPS fraction of $12\%$, despite these galaxies not having citizen science morphologies consistent with known RPS galaxies. This suggests that while using citizen science classifications to help in searching for new RPS candidates can increase the purity or efficiency when creating a new RPS sample, any preselection may miss other potential RPS candidates, potentially reducing completeness.

We also see an increase in the number of strong RPS candidates with 5\% of the GZ-selected samples, compared to 2\% for the non-GZ controls. However, due to the small numbers of galaxies in this sample, is not statistically significant.

We see the GZ-field sample has an RPS fraction of $6 \pm 1\%$, significantly lower than the GZ selected sample ($>4\sigma$ confidence interval difference) and the non-GZ selected controls ($>2\sigma$ difference). We see that non-cluster galaxies are significantly less likely to have observed features consistent with ram pressure stripping. We also see a drop in the number of strong RPS candidates compared with the GZ-selected sample. Interestingly, we see a similar value ($3\pm 1\%$) compared with the controls. However, the low numbers mean any difference is not statistically significant.

Previous studies that have performed visual searches for ram pressure stripping galaxies vary in the RPS fraction, which can depend on several factors. When considering all galaxies in a group or cluster above log(sSFR) $>-11$, the RPS fraction is typically $2-5\%$ \citep[e.g.,][]{Poggianti:2016a, Roberts:2021b, Kolcu:2022a}, which also depends on the mass of the groups and clusters considered. However, the RPS fraction has been found to be as high as $\sim16\%$  when considering all H$\alpha$ emitting galaxies in a merging cluster system \citep[e.g.,][]{Roman-Oliveira:2021a}.

Given our RPS fraction of $19\%$ without a cut in star formation rate, our results show that using the Galaxy Zoo morphologies can yield a higher fraction of potential RPS candidates in clusters than using only a colour or SFR preselection. Our fraction of RPS galaxies in the GZ-selected sample is also similar to that seen in \citet{Vulcani:2022a}, who remove passive galaxies, interacting galaxies, and non-spirals from their sample. This suggests that using using morphological criteria such as we have, as well as cuts in star formation rate or colour, will likely increase the fraction of galaxies detected with RPS features. 

Our results are similar to that of \citet{McPartland:2016a}, who identified potential RPS candidates using a combination of morphological parameters, including concentration, asymmetry, and M$_{20}$. From a morphologically reduced sample of 1263 galaxies, they find 211 potential stripping cases from expert classification, with an RPS fraction of $\sim17\%$. However, only three of these galaxies were considered new strong `jellyfish' candidates, with a further 103 considered likely candidates. While RPS galaxies are found to have unique morphological parameters \citep[see also][]{Roberts:2020a}, it is difficult to isolate RPS galaxies solely using these parameters. Incorporating multiple methods, including automated and citizen science morphologies, will likely increase the efficiency of finding new RPS candidates.

We note that these RPS fractions use different sources of imaging, with different depths and instruments. Additionally, RPS fractions use different populations as a baseline for comparison, sometimes comparing to star forming galaxies, blue cluster galaxies, or all cluster galaxies, \citep[see e.g.,][for further discussion]{Kolcu:2022a}. Comparing the fraction of RPS galaxies found between studies will often depend on these factors, as well as the subjective interpretations of visual RPS features in galaxies. If a study uses a more permissive definition of ram pressure stripping, then this would artificially increase the RPS fraction. The RPS fraction of $6\%$ in the GZ-field sample demonstrates that our expert classifications may contain false positives, but this is difficult to compare across studies. While our sample has a higher fraction of RPS candidates than many other studies, we cannot attribute all of this increase to the use of GZ morphologies.

\subsubsection{Morphological features in the expert classified RPS candidates}
\label{sec:properties_cands}
With our sample of expert classified RPS candidates defined, we can investigate the properties of these galaxies, based on the feedback of the expert classifiers.

Expert classifiers were asked what RPS features are visible in the galaxy. These include features such as tails, star formation from potential gas compressions, and asymmetries in the spiral structure (see Table \ref{tab:GASP_questions} for details). We find that most of the expert classified RPS candidates have at least two features present, with many having three or four RPS features. We do not see any difference across the four samples, suggesting that citizen scientists do not preferentially pick out galaxies with more or less features. This is true for both the entire new candidate sample as well as new strong candidates, which suggests that the strength of ram pressure stripping in these galaxies isn't necessarily correlated with the number of different types of features seen for these samples. 

However, we find that fewer classifiers were able to see a visible tail in the field sample. The fraction of expert classifiers who could determine a tail or direction of ram pressure stripping was similar across the three cluster samples (an average of 73\% could determine a tail), but was lower in the GZ-field sample (48\%). This is also true for the strong candidates across the three cluster samples (83\%), compared with (54\%) for the GZ-field sample. 

We further investigate the features seen in the new sample of ram pressure stripping candidates. We find that the GZ-selected candidates have a slight excess in the fraction of expert classified RPS candidates with tails. Conversely, the GZ-field candidates have a much lower fraction of tail features. 

Additionally, we find a higher fraction of unwinding features in the GZ-field sample. This may be due to the unwinding pattern not being a specific feature of ram pressure stripping. However, these trends are subject to large unertainties, and do not have a. high significance. However, further work may show that tails are a prominent feature that can be often identified and flagged by citizen scientists, and are more commonly associated with cluster specific processes. By comparison, an unwinding spiral arm pattern appears to be common with the field sample, and may not be a good tracer of ram pressure stripping.

\section{Discussion}
\label{sec:discussion}
\subsection{Ram pressure stripping in citizen science}
In Sect.\ \ref{sec:morphologies} we compared the morphological properties of known ram pressure stripping candidate galaxies with magnitude matched cluster and field galaxies. The results show that galaxies undergoing ram pressure stripping are visually different when assessed by citizen scientists, compared with other galaxies of similar brightness. The known RPS candidates had a higher fraction of disks, as well as more `odd' visual features compared to both cluster and field counterparts. Despite the shallow SDSS imaging data presented to each classifier, citizen scientists were able to identify morphological features that were inconsistent with the general galaxy population. 

We saw in Fig. \ref{fig:GZ_odd_breakdown_highJ} that these effects were most pronounced in galaxies with strong optical RPS features. While this fact may seem obvious, the correlation of ram pressure stripping feature strength, and the fraction of odd votes shows that the classifiers were likely picking out the RPS features. This was further confirmed in Sect.\ \ref{sec:radio_only}, where galaxies without optical RPS features had a significantly lower fraction of galaxies with $p_{\textrm{odd}} > 50\%$. Additionally, the citizen scientists were able to distinguish the ram pressure stripping features from merging and close companion systems. Panel d) of Fig. \ref{fig:GZ_multi_all} shows that very few of the ram pressure stripped candidate galaxies were classified as potential mergers.

These results demonstrate that that citizen scientists can identify rare morphological features in galaxy populations. Such examples have already been seen from results in past Galaxy Zoo studies \citep[][]{Cardamone:2009a, Keel:2012a, Keel:2022a, Zinger:2024a}. Our results demonstrate this also applies to observed galaxies with optical signatures of ram pressure stripping.

In Sect.\ \ref{sec:motivated} we use these results to motivate a new sample of ram pressure stripped candidates. When classifying images of cluster galaxies, we find that galaxies with galaxy zoo flags such as `odd' and `irregular' were more likely to have ram pressure stripping features than other cluster galaxies. We find that galaxies which are considered to be `odd', with odd features such as `irregular' and `disturbed' morphologies are the best tracers of potential ram pressure stripping. New ram pressure stripped candidates could be more easily identified by selecting galaxies based on these and other citizen science morphological parameters.
 
All of our results have used an indirect method to infer that citizen scientists can identify ram pressure stripping. In the GZ2 questionnaire, there are no questions regarding tail features, or debris that is consistent with ram pressure stripping. Our results have used indirect tracers, including a galaxy being odd, and having loosely wound spiral arms as proxies based on known ram pressure stripped galaxies. We discuss the limitations of this further in Sect.\ \ref{sec:Indirect}.

\subsection{Unwinding spirals}
\label{sec:unwind}
In Sect.\ \ref{sec:morphologies} we found that the ram pressure stripped galaxies have a significantly lower than average arm winding compared to either of the comparison samples. This result is interesting, as it matches results from \citet{Bellhouse:2021a} that showed that the H$\alpha$ emission in spiral arms of ram pressure stripped galaxies may appear to be `unwinding' as the galaxy infalls. Such galaxies have since been subject to further investigation in \citet{Vulcani:2022a}, who found example candidate unwinding galaxies to have similar properties to other ram pressure stripping galaxies when controlling for mass.

It is very possible that this effect is being seen in our results, whereby the known ram pressure stripped galaxies have been experiencing this `unwinding' or broadening of the spiral arms as they infall. However, this is not the only scenario that can explain our result. When inspecting the SDSS images of some of the loosely-wound spiral galaxies, we find that it is possible that classifiers have interpreted some of the debris as a separate spiral arm. This would give the illusion of a more loose spiral arm winding pattern. Additionally, the GZ-field sample described in Sect. \ref{sec:false} reported a higher fraction of unwinding features when classified by experts, albeit with a large uncertainty. So while the result from Sect.\ \ref{sec:morphologies} may match the scenario described in \citet{Bellhouse:2021a}, we cannot rule out biases in the classifications.

\begin{figure}
	\resizebox{\hsize}{!}{\includegraphics{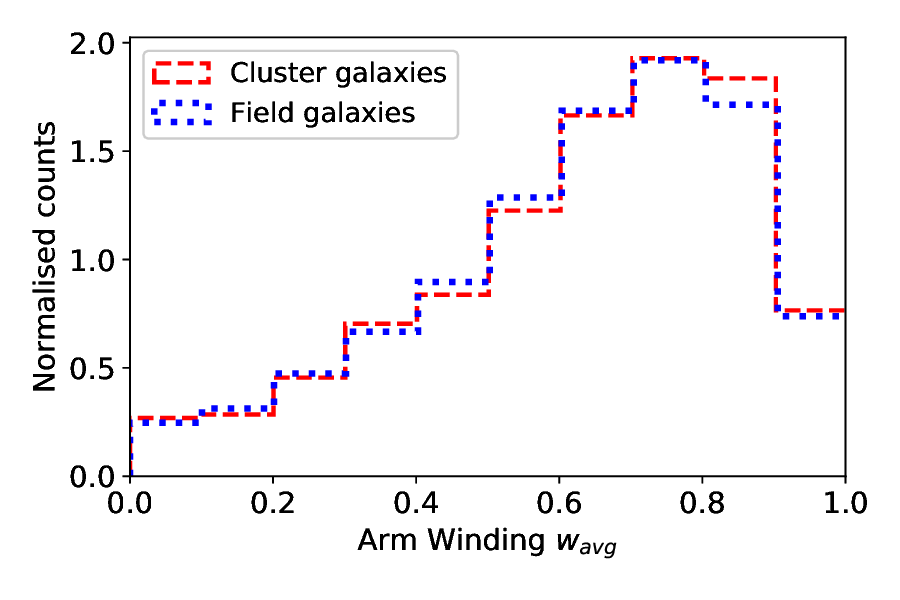}}
	    \caption{Histogram comparing the $w_{avg}$ for cluster and field face-on spiral galaxies. We do not see a difference in the fraction of loosely wound spiral arms (low $w_{avg}$ values) in clusters compared to the field, when considering all field and cluster galaxies.}
    \label{fig:arm:winding_hist}
\end{figure}

We further investigate the prevalence of these unwinding spiral galaxies in clusters by comparing the incidence of these unwinding spiral arm patterns in our cluster and field Galaxy Zoo samples. We take all the cluster and field galaxies in GZ2 using the definitions in Sect.\ \ref{sec:comparisons}, and select from these samples galaxies that are face-on and have visible spiral arms. This yields 1928 face-on cluster spirals and 18380 field spirals. When comparing the $w_{avg}$ from equation \ref{equ:armwind} for these samples, we find a very similar distribution of arm winding in both samples. Figure \ref{fig:arm:winding_hist} shows the distributions of $w_{avg}$ for both the cluster and field spiral samples. We see the distributions are closely related, with a K-S test unable to distinguish these samples at a 1$\sigma$ level. We additionally see the fraction of galaxies with $w_{avg} < 0.5$ is $\sim26\%$ in both the cluster and field sample. This would suggest that there isn't an excess of loosely wound spirals in cluster environments. 

Finally, the study of \citet{Casteels:2013a} found a link between loose winding arms in spiral galaxies and tidal pair interactions, highlighting another mechanism which is known to cause loosely wound spiral arms. These results combined suggest that the unwinding morphologies seen in many ram pressure stripped galaxies may not be a ram pressure specific phenomenon. Instead, it is likely that several processes, including ram pressure stripping, and tidal interactions are able to cause this deformation.

Interestingly, we find that for spirals with a loose spiral arm winding ($w_{avg} < 0.5$), there is an increase in the fraction of galaxies with $p_{\textrm{odd}} > 0.5$. Both the loose spiral arm galaxies in clusters, as well as in the field show a $\sim10\%$ increase in the fraction of galaxies with $p_{\textrm{odd}} > 0.5$ compared to their tighter wound counterparts (defined as $w_{avg} > 0.75$). Therefore, combined with the results of the GZ-field sample in Sect. \ref{sec:properties_cands}, it is likely that several processes, both present in and out of clusters, are responsible for the appearance of loosely wound spiral arms.

\subsection{Biases and caveats}
\label{sec:biases}
This study has investigated for the first time whether citizen scientists can identify galaxies experiencing RPS using existing data from GZ2, which was not intended to directly investigate ram pressure stripping. While this dataset provides some of the most comprehensive morphological analyses available, there are some limitations in applying these classifications to finding ram pressure stripping. We explore several of these here.

\subsubsection{Shallow imaging}
The imaging used in the Galaxy Zoo 2 classifications was from SDSS DR 7. This imaging is typically shallower than the imaging used to identify the features seen in the known ram pressure stripped sample. While several features are visible in the Legacy Survey optical imaging, only the brightest and most obvious features are visible in the SDSS imaging. We expect that this would make our results an underestimate of the fraction of odd/disturbed features seen in the ram pressure stripped galaxy sample. Recent citizen science surveys using deeper data such as Legacy Surveys and Hyper Suprime Cam \citep[e.g.,][]{Walmsley:2023a, Tanaka:2023a}, as well as neural network classifications \citep{Walmsley:2022a} will help overcome this issue in future citizen science projects searching for ram pressure stripped galaxies.

\subsubsection{Ram pressure stripping candidate sample} 
In this work we compiled a very heterogeneous sample of known RPS candidates for analysis. This includes many candidates that were only identified by their H\scaleto{$I$}{1.2ex} or radio continuum emission, as well as many galaxies that are not confirmed to be cluster members. Additionally, several galaxies have been flagged as potentially experiencing tidal interactions. This means while the sample is quite large, it is not necessarily a pure sample of ram pressure stripped galaxies. 

However, several of these issues have been mitigated. For example, all radio detected ram pressure stripping candidates without an optical classification were classified by two authors (JC and YJ) for any optical ram pressure stripping features (see Sect.\ \ref{sec:GZ2}). In this way, the entire known RPS sample will have been analysed for optical RPS signatures. Galaxies without optical features were also compared separately, so that conclusions from optical deformations and radio tails can be drawn separately.

A small minority of galaxies in the known RPS sample do not have the necessary spectroscopic redshifts to be confirmed as cluster members, meaning the morphological deformations may not be due to RPS if they are foreground or background galaxies. However, this study only requires that experts see possible ram pressure stripping, which can be compared with what is seen by citizen scientists. It is only required that these galaxies have morphological features consistent with ram pressure stripping, as determined by expert classifications, to compare with citizen science classifications. There have been cases of galaxies with features that appear like ram pressure stripping, but are found to be due to other processes \citep[e.g.,][]{Moretti:2018a, Vulcani:2018a, Vulcani:2021a}. A more detailed analysis, incorporating spatially resolved H$\alpha$ emission, kinematics, and/or H\scaleto{$I$}{1.2ex} gas is likely required to confirm if RPS is the dominant cause of any morphological disturbance \citep[but may not always confirm a single scenario, see discussions by e.g.,][]{Lee-Waddell:2018a, Serra:2023a}. 

\subsubsection{Tidal interactions and the limitations of visual classifications}
\label{sec:tidal}
In this study we have visually identified morphological features such as tails, disk compressions, and other deformations that are consistent with ram pressure stripping. Studies from the GASP survey find that  ram pressure stripping candidates identified using broad band optical morphologies are confirmed in over $80\%$ of cases using spatially resolved H$\alpha$ \citep[see discussions in][]{Lourenco:2023a}. However, there are several examples of galaxies where the deformations are more likely caused by tidal interactions \citep[e.g.][]{Vulcani:2018a, Serra:2023a}.

In Sect.\ \ref{sec:GZmorphologies}, all galaxies included in the known RPS sample were previously considered to be consistent with RPS by their respective studies. We note that this includes some candidates with ambiguous features within this sample (e.g., GMP4060, GMP4555, NGC 3860b). We choose to include these galaxies in the sample, but note that the results from Sect.\ \ref{sec:GZmorphologies} remain the same if we exclude these galaxies.

In Sect.\ \ref{sec:GZ_motivated}, galaxies were classified by expert astronomers to determine whether ram pressure stripping was present. In addition to classifying each galaxy as experiencing RPS, the expert classifiers could categorise a galaxies as experiencing a merger or tidal interaction. We can use these classifications to define a sample of tidally disturbed galaxies, analogous to Sect. \ref{sec:new_cands}, in order to measure the contribution from tidally disturbed galaxies in citizen science created samples. We define tidal candidates from the expert classified galaxies as:
\begin{itemize}
    \item f$_{\textrm{tidal}} > 0.5$ OR
    \item f$_{\textrm{RPS/tidal/unsure}} > 0.75$
\end{itemize}

\begin{figure}
	\resizebox{\hsize}{!}{\includegraphics{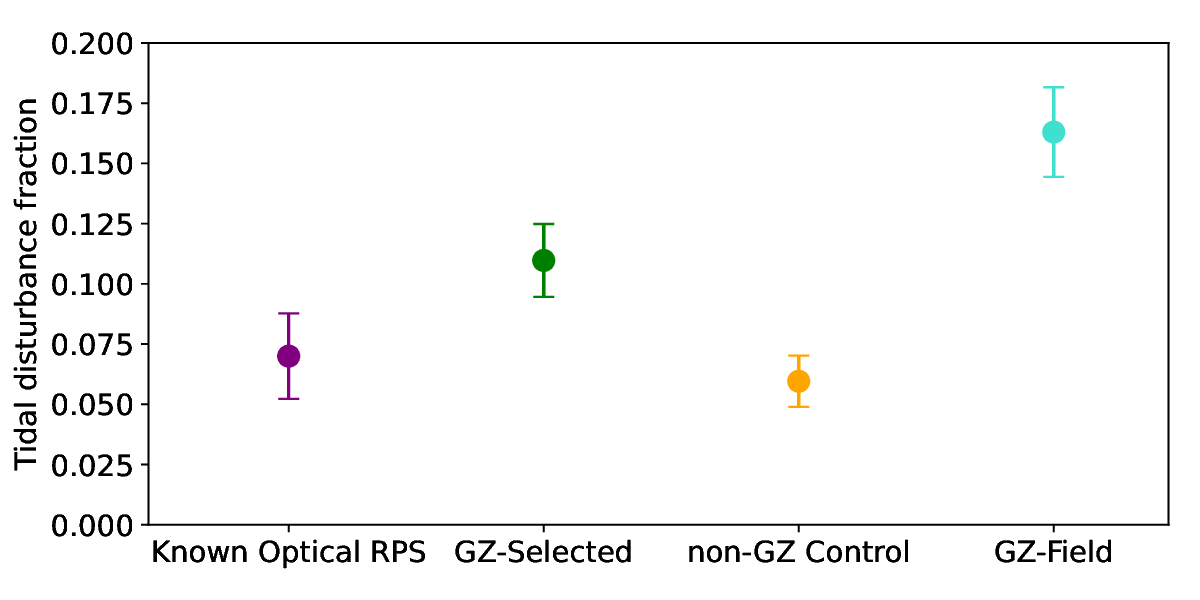}}
    \caption{Fraction of galaxies considered to be expert classified tidal candidates in each of the four samples defined in Sect.\ \ref{sec:tidal}. The GZ-selected sample has a slightly higher tidal fraction compared to the control sample, with the GZ-field sample having a higher fraction than all other samples.}
    \label{fig:GASP_merge_frac}
\end{figure}

Using this, we calculate the fraction of galaxies that experts classified as experiencing a tidal interaction. Figure \ref{fig:GASP_merge_frac} shows the fraction of tidal galaxies in each subsample. We see that approximately $8\%$ of cluster galaxies have tidal features in our samples, which increases to $16\%$ in our GZ-field sample. This suggests that when using GZ2 flags such as disturbed or irregular, approximately 8\% of galaxies are likely experiencing a tidal interaction. We find this trend is seen even with changes to the  f$_{\textrm{tidal}}$ or f$_{\textrm{RPS/tidal/unsure}}$ values (similar to Sect. \ref{sec:Selections}).

This figure shows that the use of citizen science classifications will not give a pure sample of ram pressure stripped galaxies, and when inspected by experts, will have contamination from tidal interactions. Using GZ2 classifications to select galaxies will result in an increase in the fraction of RPS candidates (see Fig. \ref{fig:GASP_rps_frac}), but will also give a slightly higher fraction of tidally disturbed galaxies as well. This is despite selecting galaxies with GZ2 classifications consistent with a non-merger. While this method improves the incidence of ram pressure stripping, gravitational interactions will also be selected using this method, and need to be accounted for. Including additional information \citep[such as a pair separation or close companions][]{Ellison:2010a, Casteels:2013a, Patton:2016a} may help in further distinguishing asymmetric features as caused by tidal interactions.

\subsubsection{Colour-matched comparisons}
\label{sec:col-matched}
In Sect.\ \ref{sec:comparisons} we generated magnitude and redshift matched comparison galaxies to compare with the known ram pressure stripped galaxies. We noted that the ram pressure stripped galaxies had significantly bluer $(g-i)$ colours compared with both the field and the cluster comparison sample. While this was done to remain agnostic as to the star formation properties of the comparison galaxies, a high number of red elliptical galaxies may not be seen as a fair comparison sample.

We therefore repeat the analysis in Sect.\ \ref{sec:GZmorphologies}, with a comparison sample designed to also match in $(g-i)$ colour. For each known RPS galaxy, we select five matched field comparison galaxies within $|\Delta M_{r}| < 0.07$, $|\Delta z| < 0.05$ and $|\Delta (g-i)| < 0.06$. We also create five matched cluster comparison galaxies within $|\Delta M_{r}| < 0.45$, $|\Delta z| < 0.135$ and $|\Delta (g-i)| < 0.12$, as per Sect.\ \ref{sec:comparisons}. 

When we compare the GZ2 morphologies with colour matched comparisons, we find that the known RPS galaxies no longer have an excess in the disk vote fraction, with $p_{\textrm{disk}} = 72\%$ in the known RPS sample, compared with $p_{\textrm{disk}} = 66\%$ for the two comparison samples. This is true both considering the whole known RPS sample, and only the strong optical RPS galaxies. The strong RPS sample has $p_{\textrm{disk}} = 78\%$, and the group and field comparison samples have  $p_{\textrm{disk}} = 73\%$ and $p_{\textrm{disk}} = 74\%$  respectively. This is likely because galaxy optical colour and morphology are known to correlate, such that selecting only blue galaxies will likely remove mostly non-disk dominated galaxies from being included in potential comparison samples.

However, the known RPS galaxies, as well as the strong optical known RPS galaxies still show a significant difference in many parameters, with a higher fraction of odd objects, with the full RPS sample having $p_{\textrm{odd}} = 23\%$ compared with $p_{\textrm{odd}} = 16\%$ in the group comparison sample, and $p_{\textrm{odd}}=53\%$ for strong RPS galaxies, compared with the group comparison fraction of $p_{\textrm{odd}} = 26\%$. We find the average arm winding score w$_{avg}$ is  0.07 lower for the known RPS galaxies compared with the group controls for both the whole sample and for strong RPS. We conclude that while ram pressure stripped galaxies may not be more likely to contain a disk than blue field or cluster comparisons, they are still more likely to have odd features and loose spiral arms as seen by GZ classifiers.

\subsubsection{Expert classified RPS candidate selection}
\label{sec:Selections}
In Sect. \ref{sec:GZ_motivated} we created a new sample of galaxies using GZ2 classifications, to test whether this can increase the fraction of RPS galaxies using expert classifications. Both the selection of the GZ-selected sample in Sect.\ \ref{sec:motivated}, and the subsequent selection of expert classified RPS candidates in Sect.\ \ref{sec:new_cands} were completed using arbitrary vote fractions.

The GZ-selected sample was created using generous debiased vote fractions in order to maximise the sample size, and thus number of potential new candidates. To ensure our results do not depend on this selection, we rerun the analysis with a more restrictive sample of GZ-selected candidates. We find the difference between the samples persists when changing the selection, and thus do not believe that the specific values chosen for the GZ-sample are a major influence in our results.

Additionally, in Sect.\ \ref{sec:new_cands}, the expert classified RPS candidates were selected with a combination of  f$_{\textrm{RPS}}$ and f$_{\textrm{RPS/tidal/unsure}}$ vote fractions. The values chosen in this instance were intended to be quite conservative, and only consider galaxies which would be considered genuine RPS candidates for use in future studies. 
To ensure this selection does not affect our results, we recalculate our RPS fraction considering a more generous f$_{\textrm{RPS}}$ and f$_{\textrm{RPS/tidal/unsure}}$ value (0.4 and 0.65), as well as a more restrictive f$_{\textrm{RPS}}$ and f$_{\textrm{RPS/tidal/unsure}}$ (0.6 and 0.85 respectively). We find that a more generous selection results in an increased RPS fraction across all samples, and a more restrictive selection results in a decreased RPS fraction. However, the difference between the GZ-selected sample and the non-GZ controls is present in both cases. Therefore, we do not consider the choice of f$_{\textrm{RPS}}$ and f$_{\textrm{RPS/tidal/unsure}}$ values to be biasing our results.

\subsubsection{Indirect classifications}
\label{sec:Indirect}
The sample used in Sect.\ \ref{sec:samp} is comprised of galaxies that have been been visually classified in previous studies to have visible optical features that are consistent with ram pressure stripping (see Table \ref{tab:GZ_jf_sources}). This gives us a reasonable confidence that any dominant morphological features seen by classifiers would be due to the ram pressure stripping features. However, as there are no questions in the Galaxy Zoo 2 questionnaire that ask specifically about features such as tails or one-sided debris, we cannot know for sure that the features seen by citizen scientist classifiers are the same ones seen in previous studies. Instead, other morphological features of the galaxy may have been seen by citizen scientists.

While we assume that this scenario is unlikely across the whole sample, we cannot estimate how frequently this may occur. Future projects from the Galaxy Zoo collaboration include updated questions about morphological disturbances, including both `major disturbance' and `minor disturbance' options. These new questions could also be useful isolate RPS features. However, we rely on the assumption that a classifier is labelling a ram pressure stripped galaxy as odd due to RPS features, and not something else.

The current Galaxy Zoo projects include a forum for classifiers to add additional comments where necessary. A search for the term `\#jellyfish' yields many potential subjects, including several known ram pressure stripped galaxies from \citet{Poggianti:2016a}, as well as many new  RPS candidates with spectacular features. This shows that classifiers can recognise jellyfish-like features, and categorise galaxies accordingly. The feedback through the Zooniverse talk forum could allow for additional RPS candidates to be found, and studied to further our knowledge of ram pressure stripping. We will present many of these candidates in a forthcoming publication.

As seen by the \#jellyfish comments, it is not unreasonable for citizen scientists to identify ram pressure stripping features directly. Asking if a classifier is able to see features such as a tail, or debris on one side of the galaxy is likely a more efficient route to finding RPS candidates with citizen science. This is currently being undertaken by several citizen science projects specifically attempting to directly find ram pressure stripping candidates. \citet{Zinger:2024a} have used the Zooniverse platform to find simulated jellyfish galaxies in imaging from illustris TNG50 and TNG100. Additionally, Bellhouse et al.\ (in prep.) is using the same platform to search for  ram pressure stripping in observational legacy survey data\footnote{\url{https://www.zooniverse.org/projects/cbellhouse/fishing-for-jellyfish-galaxies}}.

\section{Conclusion}
\label{sec:conclusion}
In this study we have investigated whether citizen science projects can identify the unique morphologies seen in galaxies undergoing ram pressure stripping. We analyse Galaxy Zoo 2 morphological classifications for a large sample of known RPS candidates and compare these to control samples. Our findings are as follows:
\begin{itemize}
    \item Known RPS candidates have morphological properties that are significantly divergent from similar galaxies in either cluster or field environments. This is seen through a higher disk fraction, and a higher fraction of containing odd features. 
    \item The odd features seen in these known RPS candidates are generally consistent with a disturbed or irregular morphology, most prevalent in `strong' ram pressure stripping candidates.
    \item $35\%$ of our sample contain no optical deformations, despite radio features consistent with RPS. These galaxies also do not contain an excess of odd features compared to the comparison samples.
    \item We find that known RPS galaxies have a lower arm winding score compared with field or cluster comparisons. This may provide evidence of the unwinding phenomenon seen in \citet{Bellhouse:2021a} and \citet{Vulcani:2022a}, although it may also be due to confusion between RPS features and spiral arms in citizen science classifications.
    \item Selecting cluster galaxies with certain Galaxy Zoo 2 classifications yields a higher fraction of ram pressure stripped galaxies ($19\%$) than other disk galaxies in clusters ($12\%$). 
    \item Using a sample on non-cluster galaxies, we find a false-positive RPS fraction of $6\%$. While significantly smaller than the two cluster samples, this highlights the difficulties in using broad band imaging to create a pure sample of RPS galaxies.
\end{itemize}

These results together suggest that citizen scientists are able to recognise RPS features in galaxies, and classify them accordingly. This is potentially despite a lack of knowledge of RPS, or the use of deep imaging often required to best identify these structures. The use of these classifications has already allowed the identification of 101 new candidates. However, care should be taken to account for tidal interactions, which can also have similar features in citizen science classifications.

Our works shows that citizen science projects such as Galaxy Zoo can be used to aid in the search for new RPS candidates. This is already being investigated in a follow up work, with an ongoing public Galaxy Zoo search for new RPS candidates underway\footnotemark[\value{footnote}]. Future wide field imaging surveys, such as LSST, may be able to employ such citizen science techniques to allow the study of cluster infall and ram pressure stripping in ways that are as yet not possible due to low numbers of galaxies found to be experiencing visible ram pressure stripping.

\section{Data Availability}
A table containing the SDSS names, sky positions, and $\overline{J_{class}}$ of all new RPS candidate galaxies found in this study is available in electronic form at the CDS via anonymous ftp to cdsarc.u-strasbg.fr (130.79.128.5) or via http://cdsweb.u-strasbg.fr/cgi-bin/qcat?J/A+A/.

\begin{acknowledgements}
      The authors thank the referee for their comments that enhanced the quality of manuscript. The authors thank Bianca Poggianti for helpful discussions and classifications for this work. The authors also wish to thank Alessandro Omizzolo, Matilde Mingozzi, Paula Calder\'on-Castillo, Koshy George, and Stephanie Tonnesen for help with expert classifications. 
JPC acknowledges financial support from ANID through FONDECYT Postdoctorado Project 3210709, as well as support from Comit\'{e} Mixto ESO-Gobierno de Chile. 
YLJ acknowledges support from the Agencia Nacional de Investigaci\'{o}n y Desarrollo (ANID) through Basal project FB210003, FONDECYT Regular projects 1241426 and 123044, and Millennium Science Initiative Program NCN2024\_112.
SLM acknowledges support from the Science and Technology Facilities Council through grant number ST/N021702/1.
KK acknowledges full financial support from ANID through FONDECYT Postdoctrorado Project 3200139. 
ACCL thanks the financial support of the National Agency for research and Development (ANID)/Scholarship Program/DOCTORADO BECAS CHILE/2019-21190049.

This publication has been made possible by the participation of hundreds of thousands of volunteers in the galaxy Zoo project on Zooniverse.org. The efforts from each volunteer is acknowledged at \url{https://authors.galaxyzoo.org/authors.html}.

Funding for the SDSS and SDSS-II has been provided by the Alfred P. Sloan Foundation, the Participating Institutions, the National Science Foundation, the U.S. Department of Energy, the National Aeronautics and Space Administration, the Japanese Monbukagakusho, the Max Planck Society, and the Higher Education Funding Council for England. The SDSS Web Site is \url{http://www.sdss.org/}.

The SDSS is managed by the Astrophysical Research Consortium for the Participating Institutions. The Participating Institutions are the American Museum of Natural History, Astrophysical Institute Potsdam, University of Basel, University of Cambridge, Case Western Reserve University, University of Chicago, Drexel University, Fermilab, the Institute for Advanced Study, the Japan Participation Group, Johns Hopkins University, the Joint Institute for Nuclear Astrophysics, the Kavli Institute for Particle Astrophysics and Cosmology, the Korean Scientist Group, the Chinese Academy of Sciences (LAMOST), Los Alamos National Laboratory, the Max-Planck-Institute for Astronomy (MPIA), the Max-Planck-Institute for Astrophysics (MPA), New Mexico State University, Ohio State University, University of Pittsburgh, University of Portsmouth, Princeton University, the United States Naval Observatory, and the University of Washington.
\end{acknowledgements}


\bibliographystyle{aa} 
\bibliography{Galzoo_refs.bib} 

\begin{thebibliography}{112}
\expandafter\ifx\csname natexlab\endcsname\relax\def\natexlab#1{#1}\fi

\bibitem[{{Abazajian} {et~al.}(2009){Abazajian}, {Adelman-McCarthy}, {Ag{\"u}eros}, {Allam}, {Allende Prieto}, {An}, {Anderson}, {Anderson}, {Annis}, {Bahcall}, {Bailer-Jones}, {Barentine}, {Bassett}, {Becker}, {Beers}, {Bell}, {Belokurov}, {Berlind}, {Berman}, {Bernardi}, {Bickerton}, {Bizyaev}, {Blakeslee}, {Blanton}, {Bochanski}, {Boroski}, {Brewington}, {Brinchmann}, {Brinkmann}, {Brunner}, {Budav{\'a}ri}, {Carey}, {Carliles}, {Carr}, {Castander}, {Cinabro}, {Connolly}, {Csabai}, {Cunha}, {Czarapata}, {Davenport}, {de Haas}, {Dilday}, {Doi}, {Eisenstein}, {Evans}, {Evans}, {Fan}, {Friedman}, {Frieman}, {Fukugita}, {G{\"a}nsicke}, {Gates}, {Gillespie}, {Gilmore}, {Gonzalez}, {Gonzalez}, {Grebel}, {Gunn}, {Gy{\"o}ry}, {Hall}, {Harding}, {Harris}, {Harvanek}, {Hawley}, {Hayes}, {Heckman}, {Hendry}, {Hennessy}, {Hindsley}, {Hoblitt}, {Hogan}, {Hogg}, {Holtzman}, {Hyde}, {Ichikawa}, {Ichikawa}, {Im}, {Ivezi{\'c}}, {Jester}, {Jiang}, {Johnson}, {Jorgensen}, {Juri{\'c}}, {Kent}, {Kessler}, {Kleinman}, {Knapp},
  {Konishi}, {Kron}, {Krzesinski}, {Kuropatkin}, {Lampeitl}, {Lebedeva}, {Lee}, {Lee}, {French Leger}, {L{\'e}pine}, {Li}, {Lima}, {Lin}, {Long}, {Loomis}, {Loveday}, {Lupton}, {Magnier}, {Malanushenko}, {Malanushenko}, {Mandelbaum}, {Margon}, {Marriner}, {Mart{\'\i}nez-Delgado}, {Matsubara}, {McGehee}, {McKay}, {Meiksin}, {Morrison}, {Mullally}, {Munn}, {Murphy}, {Nash}, {Nebot}, {Neilsen}, {Newberg}, {Newman}, {Nichol}, {Nicinski}, {Nieto-Santisteban}, {Nitta}, {Okamura}, {Oravetz}, {Ostriker}, {Owen}, {Padmanabhan}, {Pan}, {Park}, {Pauls}, {Peoples}, {Percival}, {Pier}, {Pope}, {Pourbaix}, {Price}, {Purger}, {Quinn}, {Raddick}, {Re Fiorentin}, {Richards}, {Richmond}, {Riess}, {Rix}, {Rockosi}, {Sako}, {Schlegel}, {Schneider}, {Scholz}, {Schreiber}, {Schwope}, {Seljak}, {Sesar}, {Sheldon}, {Shimasaku}, {Sibley}, {Simmons}, {Sivarani}, {Allyn Smith}, {Smith}, {Smol{\v{c}}i{\'c}}, {Snedden}, {Stebbins}, {Steinmetz}, {Stoughton}, {Strauss}, {SubbaRao}, {Suto}, {Szalay}, {Szapudi}, {Szkody}, {Tanaka},
  {Tegmark}, {Teodoro}, {Thakar}, {Tremonti}, {Tucker}, {Uomoto}, {Vanden Berk}, {Vandenberg}, {Vidrih}, {Vogeley}, {Voges}, {Vogt}, {Wadadekar}, {Watters}, {Weinberg}, {West}, {White}, {Wilhite}, {Wonders}, {Yanny}, {Yocum}, {York}, {Zehavi}, {Zibetti}, \& {Zucker}}]{Abazajian:2009a}
{Abazajian}, K.~N., {Adelman-McCarthy}, J.~K., {Ag{\"u}eros}, M.~A., {et~al.} 2009, \apjs, 182, 543

\bibitem[{{Abramson} {et~al.}(2011){Abramson}, {Kenney}, {Crowl}, {Chung}, {van Gorkom}, {Vollmer}, \& {Schiminovich}}]{Abramson:2011a}
{Abramson}, A., {Kenney}, J. D.~P., {Crowl}, H.~H., {et~al.} 2011, \aj, 141, 164

\bibitem[{{Adelman-McCarthy} {et~al.}(2008){Adelman-McCarthy}, {Ag{\"u}eros}, {Allam}, {Allende Prieto}, {Anderson}, {Anderson}, {Annis}, {Bahcall}, {Bailer-Jones}, {Baldry}, {Barentine}, {Bassett}, {Becker}, {Beers}, {Bell}, {Berlind}, {Bernardi}, {Blanton}, {Bochanski}, {Boroski}, {Brinchmann}, {Brinkmann}, {Brunner}, {Budav{\'a}ri}, {Carliles}, {Carr}, {Castander}, {Cinabro}, {Cool}, {Covey}, {Csabai}, {Cunha}, {Davenport}, {Dilday}, {Doi}, {Eisenstein}, {Evans}, {Fan}, {Finkbeiner}, {Friedman}, {Frieman}, {Fukugita}, {G{\"a}nsicke}, {Gates}, {Gillespie}, {Glazebrook}, {Gray}, {Grebel}, {Gunn}, {Gurbani}, {Hall}, {Harding}, {Harvanek}, {Hawley}, {Hayes}, {Heckman}, {Hendry}, {Hindsley}, {Hirata}, {Hogan}, {Hogg}, {Hyde}, {Ichikawa}, {Ivezi{\'c}}, {Jester}, {Johnson}, {Jorgensen}, {Juri{\'c}}, {Kent}, {Kessler}, {Kleinman}, {Knapp}, {Kron}, {Krzesinski}, {Kuropatkin}, {Lamb}, {Lampeitl}, {Lebedeva}, {Lee}, {French Leger}, {L{\'e}pine}, {Lima}, {Lin}, {Long}, {Loomis}, {Loveday}, {Lupton}, {Malanushenko},
  {Malanushenko}, {Mandelbaum}, {Margon}, {Marriner}, {Mart{\'\i}nez-Delgado}, {Matsubara}, {McGehee}, {McKay}, {Meiksin}, {Morrison}, {Munn}, {Nakajima}, {Neilsen}, {Newberg}, {Nichol}, {Nicinski}, {Nieto-Santisteban}, {Nitta}, {Okamura}, {Owen}, {Oyaizu}, {Padmanabhan}, {Pan}, {Park}, {Peoples}, {Pier}, {Pope}, {Purger}, {Raddick}, {Re Fiorentin}, {Richards}, {Richmond}, {Riess}, {Rix}, {Rockosi}, {Sako}, {Schlegel}, {Schneider}, {Schreiber}, {Schwope}, {Seljak}, {Sesar}, {Sheldon}, {Shimasaku}, {Sivarani}, {Allyn Smith}, {Snedden}, {Steinmetz}, {Strauss}, {SubbaRao}, {Suto}, {Szalay}, {Szapudi}, {Szkody}, {Tegmark}, {Thakar}, {Tremonti}, {Tucker}, {Uomoto}, {Vanden Berk}, {Vandenberg}, {Vidrih}, {Vogeley}, {Voges}, {Vogt}, {Wadadekar}, {Weinberg}, {West}, {White}, {Wilhite}, {Yanny}, {Yocum}, {York}, {Zehavi}, \& {Zucker}}]{Adelman-McCarthy:2008a}
{Adelman-McCarthy}, J.~K., {Ag{\"u}eros}, M.~A., {Allam}, S.~S., {et~al.} 2008, \apjs, 175, 297

\bibitem[{{Bamford} {et~al.}(2009){Bamford}, {Nichol}, {Baldry}, {Land}, {Lintott}, {Schawinski}, {Slosar}, {Szalay}, {Thomas}, {Torki}, {Andreescu}, {Edmondson}, {Miller}, {Murray}, {Raddick}, \& {Vandenberg}}]{Bamford:2009a}
{Bamford}, S.~P., {Nichol}, R.~C., {Baldry}, I.~K., {et~al.} 2009, \mnras, 393, 1324

\bibitem[{{Bekki}(2009)}]{Bekki:2009a}
{Bekki}, K. 2009, \mnras, 399, 2221

\bibitem[{{Bekki}(2021)}]{Bekki:2021a}
{Bekki}, K. 2021, \aap, 647, A120

\bibitem[{{Bekki} \& {Couch}(2003)}]{Bekki:2003a}
{Bekki}, K. \& {Couch}, W.~J. 2003, \apjl, 596, L13

\bibitem[{{Bellhouse} {et~al.}(2017){Bellhouse}, {Jaff{\'e}}, {Hau}, {McGee}, {Poggianti}, {Moretti}, {Gullieuszik}, {Bettoni}, {Fasano}, {D'Onofrio}, {Fritz}, {Omizzolo}, {Sheen}, \& {Vulcani}}]{Bellhouse:2017a}
{Bellhouse}, C., {Jaff{\'e}}, Y.~L., {Hau}, G.~K.~T., {et~al.} 2017, \apj, 844, 49

\bibitem[{{Bellhouse} {et~al.}(2021){Bellhouse}, {McGee}, {Smith}, {Poggianti}, {Jaff{\'e}}, {Kraljic}, {Franchetto}, {Fritz}, {Vulcani}, {Tonnesen}, {Roediger}, {Moretti}, {Gullieuszik}, \& {Shin}}]{Bellhouse:2021a}
{Bellhouse}, C., {McGee}, S.~L., {Smith}, R., {et~al.} 2021, \mnras, 500, 1285

\bibitem[{{Bellhouse} {et~al.}(2022){Bellhouse}, {Poggianti}, {Moretti}, {Vulcani}, {Werle}, {Gullieuszik}, {Radovich}, {Jaff{\'e}}, {Fritz}, {Ignesti}, {Bacchini}, {Tomi{\v{c}}i{\'c}}, {Richard}, \& {Soucail}}]{Bellhouse:2022a}
{Bellhouse}, C., {Poggianti}, B., {Moretti}, A., {et~al.} 2022, \apj, 937, 18

\bibitem[{{Boselli} {et~al.}(2016){Boselli}, {Cuillandre}, {Fossati}, {Boissier}, {Bomans}, {Consolandi}, {Anselmi}, {Cortese}, {C{\^o}t{\'e}}, {Durrell}, {Ferrarese}, {Fumagalli}, {Gavazzi}, {Gwyn}, {Hensler}, {Sun}, \& {Toloba}}]{Boselli:2016a}
{Boselli}, A., {Cuillandre}, J.~C., {Fossati}, M., {et~al.} 2016, \aap, 587, A68

\bibitem[{{Boselli} {et~al.}(2022){Boselli}, {Fossati}, \& {Sun}}]{Boselli:2022a}
{Boselli}, A., {Fossati}, M., \& {Sun}, M. 2022, \aapr, 30, 3

\bibitem[{{Boselli} \& {Gavazzi}(2006)}]{Boselli:2006a}
{Boselli}, A. \& {Gavazzi}, G. 2006, \pasp, 118, 517

\bibitem[{{Brown} {et~al.}(2023){Brown}, {Roberts}, {Thorp}, {Ellison}, {Zabel}, {Wilson}, {Bah{\'e}}, {Bisaria}, {Bolatto}, {Boselli}, {Chung}, {Cortese}, {Catinella}, {Davis}, {Jim{\'e}nez-Donaire}, {Lagos}, {Lee}, {Parker}, {Smith}, {Spekkens}, {Stevens}, {Villanueva}, \& {Watts}}]{Brown:2023a}
{Brown}, T., {Roberts}, I.~D., {Thorp}, M., {et~al.} 2023, \apj, 956, 37

\bibitem[{{Cardamone} {et~al.}(2009){Cardamone}, {Schawinski}, {Sarzi}, {Bamford}, {Bennert}, {Urry}, {Lintott}, {Keel}, {Parejko}, {Nichol}, {Thomas}, {Andreescu}, {Murray}, {Raddick}, {Slosar}, {Szalay}, \& {Vandenberg}}]{Cardamone:2009a}
{Cardamone}, C., {Schawinski}, K., {Sarzi}, M., {et~al.} 2009, \mnras, 399, 1191

\bibitem[{{Casteels} {et~al.}(2013){Casteels}, {Bamford}, {Skibba}, {Masters}, {Lintott}, {Keel}, {Schawinski}, {Nichol}, \& {Smith}}]{Casteels:2013a}
{Casteels}, K. R.~V., {Bamford}, S.~P., {Skibba}, R.~A., {et~al.} 2013, \mnras, 429, 1051

\bibitem[{{Cattorini} {et~al.}(2023){Cattorini}, {Gavazzi}, {Boselli}, \& {Fossati}}]{Cattorini:2023a}
{Cattorini}, F., {Gavazzi}, G., {Boselli}, A., \& {Fossati}, M. 2023, \aap, 671, A118

\bibitem[{{Chung} {et~al.}(2009){Chung}, {van Gorkom}, {Kenney}, {Crowl}, \& {Vollmer}}]{Chung:2009a}
{Chung}, A., {van Gorkom}, J.~H., {Kenney}, J. D.~P., {Crowl}, H., \& {Vollmer}, B. 2009, \aj, 138, 1741

\bibitem[{{Chung} {et~al.}(2007){Chung}, {van Gorkom}, {Kenney}, \& {Vollmer}}]{Chung:2007a}
{Chung}, A., {van Gorkom}, J.~H., {Kenney}, J. D.~P., \& {Vollmer}, B. 2007, \apjl, 659, L115

\bibitem[{{Conselice}(2003)}]{Conselice:2003a}
{Conselice}, C.~J. 2003, \apjs, 147, 1

\bibitem[{{Conselice} \& {Gallagher}(1998)}]{Conselice:1998a}
{Conselice}, C.~J. \& {Gallagher}, John~S., I. 1998, \mnras, 297, L34

\bibitem[{{Cortese} {et~al.}(2021){Cortese}, {Catinella}, \& {Smith}}]{Cortese:2021a}
{Cortese}, L., {Catinella}, B., \& {Smith}, R. 2021, \pasa, 38, e035

\bibitem[{{Crowl} \& {Kenney}(2008)}]{Crowl:2008a}
{Crowl}, H.~H. \& {Kenney}, J. D.~P. 2008, \aj, 136, 1623

\bibitem[{{Crowl} {et~al.}(2005){Crowl}, {Kenney}, {van Gorkom}, \& {Vollmer}}]{Crowl:2005a}
{Crowl}, H.~H., {Kenney}, J. D.~P., {van Gorkom}, J.~H., \& {Vollmer}, B. 2005, \aj, 130, 65

\bibitem[{{Darg} {et~al.}(2010{\natexlab{a}}){Darg}, {Kaviraj}, {Lintott}, {Schawinski}, {Sarzi}, {Bamford}, {Silk}, {Andreescu}, {Murray}, {Nichol}, {Raddick}, {Slosar}, {Szalay}, {Thomas}, \& {Vandenberg}}]{Darg:2010b}
{Darg}, D.~W., {Kaviraj}, S., {Lintott}, C.~J., {et~al.} 2010{\natexlab{a}}, \mnras, 401, 1552

\bibitem[{{Darg} {et~al.}(2010{\natexlab{b}}){Darg}, {Kaviraj}, {Lintott}, {Schawinski}, {Sarzi}, {Bamford}, {Silk}, {Proctor}, {Andreescu}, {Murray}, {Nichol}, {Raddick}, {Slosar}, {Szalay}, {Thomas}, \& {Vandenberg}}]{Darg:2010a}
{Darg}, D.~W., {Kaviraj}, S., {Lintott}, C.~J., {et~al.} 2010{\natexlab{b}}, \mnras, 401, 1043

\bibitem[{{Dey} {et~al.}(2019){Dey}, {Schlegel}, {Lang}, {Blum}, {Burleigh}, {Fan}, {Findlay}, {Finkbeiner}, {Herrera}, {Juneau}, {Landriau}, {Levi}, {McGreer}, {Meisner}, {Myers}, {Moustakas}, {Nugent}, {Patej}, {Schlafly}, {Walker}, {Valdes}, {Weaver}, {Y{\`e}che}, {Zou}, {Zhou}, {Abareshi}, {Abbott}, {Abolfathi}, {Aguilera}, {Alam}, {Allen}, {Alvarez}, {Annis}, {Ansarinejad}, {Aubert}, {Beechert}, {Bell}, {BenZvi}, {Beutler}, {Bielby}, {Bolton}, {Brice{\~n}o}, {Buckley-Geer}, {Butler}, {Calamida}, {Carlberg}, {Carter}, {Casas}, {Castander}, {Choi}, {Comparat}, {Cukanovaite}, {Delubac}, {DeVries}, {Dey}, {Dhungana}, {Dickinson}, {Ding}, {Donaldson}, {Duan}, {Duckworth}, {Eftekharzadeh}, {Eisenstein}, {Etourneau}, {Fagrelius}, {Farihi}, {Fitzpatrick}, {Font-Ribera}, {Fulmer}, {G{\"a}nsicke}, {Gaztanaga}, {George}, {Gerdes}, {Gontcho}, {Gorgoni}, {Green}, {Guy}, {Harmer}, {Hernandez}, {Honscheid}, {Huang}, {James}, {Jannuzi}, {Jiang}, {Joyce}, {Karcher}, {Karkar}, {Kehoe}, {Kneib}, {Kueter-Young}, {Lan},
  {Lauer}, {Le Guillou}, {Le Van Suu}, {Lee}, {Lesser}, {Perreault Levasseur}, {Li}, {Mann}, {Marshall}, {Mart{\'\i}nez-V{\'a}zquez}, {Martini}, {du Mas des Bourboux}, {McManus}, {Meier}, {M{\'e}nard}, {Metcalfe}, {Mu{\~n}oz-Guti{\'e}rrez}, {Najita}, {Napier}, {Narayan}, {Newman}, {Nie}, {Nord}, {Norman}, {Olsen}, {Paat}, {Palanque-Delabrouille}, {Peng}, {Poppett}, {Poremba}, {Prakash}, {Rabinowitz}, {Raichoor}, {Rezaie}, {Robertson}, {Roe}, {Ross}, {Ross}, {Rudnick}, {Safonova}, {Saha}, {S{\'a}nchez}, {Savary}, {Schweiker}, {Scott}, {Seo}, {Shan}, {Silva}, {Slepian}, {Soto}, {Sprayberry}, {Staten}, {Stillman}, {Stupak}, {Summers}, {Sien Tie}, {Tirado}, {Vargas-Maga{\~n}a}, {Vivas}, {Wechsler}, {Williams}, {Yang}, {Yang}, {Yapici}, {Zaritsky}, {Zenteno}, {Zhang}, {Zhang}, {Zhou}, \& {Zhou}}]{Dey:2019a}
{Dey}, A., {Schlegel}, D.~J., {Lang}, D., {et~al.} 2019, \aj, 157, 168

\bibitem[{{Dieleman} {et~al.}(2015){Dieleman}, {Willett}, \& {Dambre}}]{Dieleman:2015a}
{Dieleman}, S., {Willett}, K.~W., \& {Dambre}, J. 2015, \mnras, 450, 1441

\bibitem[{{Dom{\'\i}nguez S{\'a}nchez} {et~al.}(2018){Dom{\'\i}nguez S{\'a}nchez}, {Huertas-Company}, {Bernardi}, {Tuccillo}, \& {Fischer}}]{Dominguez-Sanchez:2018a}
{Dom{\'\i}nguez S{\'a}nchez}, H., {Huertas-Company}, M., {Bernardi}, M., {Tuccillo}, D., \& {Fischer}, J.~L. 2018, \mnras, 476, 3661

\bibitem[{{Durret} {et~al.}(2021){Durret}, {Chiche}, {Lobo}, \& {Jauzac}}]{Durret:2021a}
{Durret}, F., {Chiche}, S., {Lobo}, C., \& {Jauzac}, M. 2021, \aap, 648, A63

\bibitem[{{Ebeling} \& {Kalita}(2019)}]{Ebeling:2019a}
{Ebeling}, H. \& {Kalita}, B.~S. 2019, \apj, 882, 127

\bibitem[{{Ellison} {et~al.}(2010){Ellison}, {Patton}, {Simard}, {McConnachie}, {Baldry}, \& {Mendel}}]{Ellison:2010a}
{Ellison}, S.~L., {Patton}, D.~R., {Simard}, L., {et~al.} 2010, \mnras, 407, 1514

\bibitem[{{Fossati} {et~al.}(2016){Fossati}, {Fumagalli}, {Boselli}, {Gavazzi}, {Sun}, \& {Wilman}}]{Fossati:2016a}
{Fossati}, M., {Fumagalli}, M., {Boselli}, A., {et~al.} 2016, \mnras, 455, 2028

\bibitem[{{Gavazzi} {et~al.}(2001){Gavazzi}, {Boselli}, {Mayer}, {Iglesias-Paramo}, {V{\'\i}lchez}, \& {Carrasco}}]{Gavazzi:2001a}
{Gavazzi}, G., {Boselli}, A., {Mayer}, L., {et~al.} 2001, \apjl, 563, L23

\bibitem[{{Gavazzi} {et~al.}(2017){Gavazzi}, {Consolandi}, {Yagi}, \& {Yoshida}}]{Gavazzi:2017a}
{Gavazzi}, G., {Consolandi}, G., {Yagi}, M., \& {Yoshida}, M. 2017, \aap, 606, A131

\bibitem[{{Gavazzi} {et~al.}(1995){Gavazzi}, {Contursi}, {Carrasco}, {Boselli}, {Kennicutt}, {Scodeggio}, \& {Jaffe}}]{Gavazzi:1995a}
{Gavazzi}, G., {Contursi}, A., {Carrasco}, L., {et~al.} 1995, \aap, 304, 325

\bibitem[{{Gavazzi} {et~al.}(1984){Gavazzi}, {Tarenghi}, {Jaffe}, {Butcher}, \& {Boksenberg}}]{Gavazzi:1984a}
{Gavazzi}, G., {Tarenghi}, M., {Jaffe}, W., {Butcher}, H., \& {Boksenberg}, A. 1984, \aap, 137, 235

\bibitem[{{George} {et~al.}(2019){George}, {Poggianti}, {Bellhouse}, {Radovich}, {Fritz}, {Paladino}, {Bettoni}, {Jaff{\'e}}, {Moretti}, {Gullieuszik}, {Vulcani}, {Fasano}, {Stalin}, {Subramaniam}, \& {Tandon}}]{George:2019a}
{George}, K., {Poggianti}, B.~M., {Bellhouse}, C., {et~al.} 2019, \mnras, 487, 3102

\bibitem[{{Gullieuszik} {et~al.}(2020){Gullieuszik}, {Poggianti}, {McGee}, {Moretti}, {Vulcani}, {Tonnesen}, {Roediger}, {Jaff{\'e}}, {Fritz}, {Franchetto}, {Omizzolo}, {Bettoni}, {Radovich}, \& {Wolter}}]{Gullieuszik:2020a}
{Gullieuszik}, M., {Poggianti}, B.~M., {McGee}, S.~L., {et~al.} 2020, \apj, 899, 13

\bibitem[{{Gunn} \& {Gott}(1972)}]{Gunn:1972a}
{Gunn}, J.~E. \& {Gott}, J.~Richard, I. 1972, \apj, 176, 1

\bibitem[{{Hart} {et~al.}(2017){Hart}, {Bamford}, {Casteels}, {Kruk}, {Lintott}, \& {Masters}}]{Hart:2017a}
{Hart}, R.~E., {Bamford}, S.~P., {Casteels}, K. R.~V., {et~al.} 2017, \mnras, 468, 1850

\bibitem[{{Hart} {et~al.}(2018){Hart}, {Bamford}, {Keel}, {Kruk}, {Masters}, {Simmons}, \& {Smethurst}}]{Hart:2018a}
{Hart}, R.~E., {Bamford}, S.~P., {Keel}, W.~C., {et~al.} 2018, \mnras, 478, 932

\bibitem[{{Hart} {et~al.}(2016){Hart}, {Bamford}, {Willett}, {Masters}, {Cardamone}, {Lintott}, {Mackay}, {Nichol}, {Rosslowe}, {Simmons}, \& {Smethurst}}]{Hart:2016a}
{Hart}, R.~E., {Bamford}, S.~P., {Willett}, K.~W., {et~al.} 2016, \mnras, 461, 3663

\bibitem[{{Holincheck} {et~al.}(2016){Holincheck}, {Wallin}, {Borne}, {Fortson}, {Lintott}, {Smith}, {Bamford}, {Keel}, \& {Parrish}}]{Holincheck:2016a}
{Holincheck}, A.~J., {Wallin}, J.~F., {Borne}, K., {et~al.} 2016, \mnras, 459, 720

\bibitem[{{Huertas-Company} {et~al.}(2015){Huertas-Company}, {Gravet}, {Cabrera-Vives}, {P{\'e}rez-Gonz{\'a}lez}, {Kartaltepe}, {Barro}, {Bernardi}, {Mei}, {Shankar}, {Dimauro}, {Bell}, {Kocevski}, {Koo}, {Faber}, \& {Mcintosh}}]{Huertas-Company:2015a}
{Huertas-Company}, M., {Gravet}, R., {Cabrera-Vives}, G., {et~al.} 2015, \apjs, 221, 8

\bibitem[{{Jaff{\'e}} {et~al.}(2018){Jaff{\'e}}, {Poggianti}, {Moretti}, {Gullieuszik}, {Smith}, {Vulcani}, {Fasano}, {Fritz}, {Tonnesen}, {Bettoni}, {Hau}, {Biviano}, {Bellhouse}, \& {McGee}}]{Jaffe:2018a}
{Jaff{\'e}}, Y.~L., {Poggianti}, B.~M., {Moretti}, A., {et~al.} 2018, \mnras, 476, 4753

\bibitem[{{Jaff{\'e}} {et~al.}(2015){Jaff{\'e}}, {Smith}, {Candlish}, {Poggianti}, {Sheen}, \& {Verheijen}}]{Jaffe:2015a}
{Jaff{\'e}}, Y.~L., {Smith}, R., {Candlish}, G.~N., {et~al.} 2015, \mnras, 448, 1715

\bibitem[{{Kapferer} {et~al.}(2009){Kapferer}, {Sluka}, {Schindler}, {Ferrari}, \& {Ziegler}}]{Kapferer:2009a}
{Kapferer}, W., {Sluka}, C., {Schindler}, S., {Ferrari}, C., \& {Ziegler}, B. 2009, \aap, 499, 87

\bibitem[{{Keel} {et~al.}(2012){Keel}, {Lintott}, {Schawinski}, {Bennert}, {Thomas}, {Manning}, {Chojnowski}, {van Arkel}, \& {Lynn}}]{Keel:2012a}
{Keel}, W.~C., {Lintott}, C.~J., {Schawinski}, K., {et~al.} 2012, \aj, 144, 66

\bibitem[{{Keel} {et~al.}(2022){Keel}, {Tate}, {Wong}, {Banfield}, {Lintott}, {Masters}, {Simmons}, {Scarlata}, {Cardamone}, {Smethurst}, {Fortson}, {Shanahan}, {Kruk}, {Garland}, {Hancock}, \& {O'Ryan}}]{Keel:2022a}
{Keel}, W.~C., {Tate}, J., {Wong}, O.~I., {et~al.} 2022, \aj, 163, 150

\bibitem[{{Kenney} {et~al.}(2015){Kenney}, {Abramson}, \& {Bravo-Alfaro}}]{Kenney:2015a}
{Kenney}, J. D.~P., {Abramson}, A., \& {Bravo-Alfaro}, H. 2015, \aj, 150, 59

\bibitem[{{Kolcu} {et~al.}(2022){Kolcu}, {Crossett}, {Bellhouse}, \& {McGee}}]{Kolcu:2022a}
{Kolcu}, T., {Crossett}, J.~P., {Bellhouse}, C., \& {McGee}, S. 2022, \mnras, 515, 5877

\bibitem[{{Krabbe} {et~al.}(2024){Krabbe}, {Hernandez-Jimenez}, {Mendes de Oliveira}, {Jaffe}, {Oliveira}, {Cardoso}, {Smith Castelli}, {Dors}, {Cortesi}, \& {Crossett}}]{Krabbe:2024a}
{Krabbe}, A.~C., {Hernandez-Jimenez}, J.~A., {Mendes de Oliveira}, C., {et~al.} 2024, \mnras, 528, 1125

\bibitem[{{Lambrides} {et~al.}(2021){Lambrides}, {Watts}, {Chiaberge}, {Tchernyshyov}, {Kirkpatrick}, {Meyer}, {Heckman}, {Simons}, {Amram}, {Hall}, {Long}, \& {Norman}}]{Lambrides:2021a}
{Lambrides}, E.~L., {Watts}, D.~J., {Chiaberge}, M., {et~al.} 2021, \apj, 919, 43

\bibitem[{{Lee-Waddell} {et~al.}(2018){Lee-Waddell}, {Serra}, {Koribalski}, {Venhola}, {Iodice}, {Catinella}, {Cortese}, {Peletier}, {Popping}, {Keenan}, \& {Capaccioli}}]{Lee-Waddell:2018a}
{Lee-Waddell}, K., {Serra}, P., {Koribalski}, B., {et~al.} 2018, \mnras, 474, 1108

\bibitem[{{Lintott} {et~al.}(2008){Lintott}, {Schawinski}, {Slosar}, {Land}, {Bamford}, {Thomas}, {Raddick}, {Nichol}, {Szalay}, {Andreescu}, {Murray}, \& {Vandenberg}}]{Lintott:2008a}
{Lintott}, C.~J., {Schawinski}, K., {Slosar}, A., {et~al.} 2008, \mnras, 389, 1179

\bibitem[{{Liu} {et~al.}(2021){Liu}, {Yee}, {Drissen}, {Sivanandam}, {Pintos-Castro}, {Alcorn}, {Hsieh}, {Lin}, {Lin}, {Muzzin}, {Noble}, \& {Old}}]{Liu:2021a}
{Liu}, Q., {Yee}, H.~K.~C., {Drissen}, L., {et~al.} 2021, \apj, 908, 228

\bibitem[{{Lotz} {et~al.}(2004){Lotz}, {Primack}, \& {Madau}}]{Lotz:2004a}
{Lotz}, J.~M., {Primack}, J., \& {Madau}, P. 2004, \aj, 128, 163

\bibitem[{{Louren{\c{c}}o} {et~al.}(2023){Louren{\c{c}}o}, {Jaff{\'e}}, {Vulcani}, {Biviano}, {Poggianti}, {Moretti}, {Kelkar}, {Crossett}, {Gitti}, {Smith}, {Lagan{\'a}}, {Gullieuszik}, {Ignesti}, {McGee}, {Wolter}, {Sonkamble}, \& {M{\"u}ller}}]{Lourenco:2023a}
{Louren{\c{c}}o}, A. C.~C., {Jaff{\'e}}, Y.~L., {Vulcani}, B., {et~al.} 2023, \mnras, 526, 4831

\bibitem[{{Masters} {et~al.}(2019){Masters}, {Lintott}, {Hart}, {Kruk}, {Smethurst}, {Casteels}, {Keel}, {Simmons}, {Stanescu}, {Tate}, \& {Tomi}}]{Masters:2019a}
{Masters}, K.~L., {Lintott}, C.~J., {Hart}, R.~E., {et~al.} 2019, \mnras, 487, 1808

\bibitem[{{McPartland} {et~al.}(2016){McPartland}, {Ebeling}, {Roediger}, \& {Blumenthal}}]{McPartland:2016a}
{McPartland}, C., {Ebeling}, H., {Roediger}, E., \& {Blumenthal}, K. 2016, \mnras, 455, 2994

\bibitem[{{Melvin} {et~al.}(2014){Melvin}, {Masters}, {Lintott}, {Nichol}, {Simmons}, {Bamford}, {Casteels}, {Cheung}, {Edmondson}, {Fortson}, {Schawinski}, {Skibba}, {Smith}, \& {Willett}}]{Melvin:2014a}
{Melvin}, T., {Masters}, K., {Lintott}, C., {et~al.} 2014, \mnras, 438, 2882

\bibitem[{{Merluzzi} {et~al.}(2013){Merluzzi}, {Busarello}, {Dopita}, {Haines}, {Steinhauser}, {Mercurio}, {Rifatto}, {Smith}, \& {Schindler}}]{Merluzzi:2013a}
{Merluzzi}, P., {Busarello}, G., {Dopita}, M.~A., {et~al.} 2013, \mnras, 429, 1747

\bibitem[{{Moretti} {et~al.}(2018){Moretti}, {Paladino}, {Poggianti}, {D'Onofrio}, {Bettoni}, {Gullieuszik}, {Jaff{\'e}}, {Vulcani}, {Fasano}, {Fritz}, \& {Torstensson}}]{Moretti:2018a}
{Moretti}, A., {Paladino}, R., {Poggianti}, B.~M., {et~al.} 2018, \mnras, 480, 2508

\bibitem[{{Moretti} {et~al.}(2020){Moretti}, {Paladino}, {Poggianti}, {Serra}, {Ramatsoku}, {Franchetto}, {Deb}, {Gullieuszik}, {Tomi{\v{c}}i{\'c}}, {Mingozzi}, {Vulcani}, {Radovich}, {Bettoni}, \& {Fritz}}]{Moretti:2020a}
{Moretti}, A., {Paladino}, R., {Poggianti}, B.~M., {et~al.} 2020, \apjl, 897, L30

\bibitem[{{M{\"u}ller} {et~al.}(2021){M{\"u}ller}, {Poggianti}, {Pfrommer}, {Adebahr}, {Serra}, {Ignesti}, {Sparre}, {Gitti}, {Dettmar}, {Vulcani}, \& {Moretti}}]{Muller:2021a}
{M{\"u}ller}, A., {Poggianti}, B.~M., {Pfrommer}, C., {et~al.} 2021, Nature Astronomy, 5, 159

\bibitem[{{Patton} {et~al.}(2016){Patton}, {Qamar}, {Ellison}, {Bluck}, {Simard}, {Mendel}, {Moreno}, \& {Torrey}}]{Patton:2016a}
{Patton}, D.~R., {Qamar}, F.~D., {Ellison}, S.~L., {et~al.} 2016, \mnras, 461, 2589

\bibitem[{{Pawlik} {et~al.}(2016){Pawlik}, {Wild}, {Walcher}, {Johansson}, {Villforth}, {Rowlands}, {Mendez-Abreu}, \& {Hewlett}}]{Pawlik:2016a}
{Pawlik}, M.~M., {Wild}, V., {Walcher}, C.~J., {et~al.} 2016, \mnras, 456, 3032

\bibitem[{{Peluso} {et~al.}(2022){Peluso}, {Vulcani}, {Poggianti}, {Moretti}, {Radovich}, {Smith}, {Jaff{\'e}}, {Crossett}, {Gullieuszik}, {Fritz}, \& {Ignesti}}]{Peluso:2022a}
{Peluso}, G., {Vulcani}, B., {Poggianti}, B.~M., {et~al.} 2022, \apj, 927, 130

\bibitem[{{Poggianti} {et~al.}(2016){Poggianti}, {Fasano}, {Omizzolo}, {Gullieuszik}, {Bettoni}, {Moretti}, {Paccagnella}, {Jaff{\'e}}, {Vulcani}, {Fritz}, {Couch}, \& {D'Onofrio}}]{Poggianti:2016a}
{Poggianti}, B.~M., {Fasano}, G., {Omizzolo}, A., {et~al.} 2016, \aj, 151, 78

\bibitem[{{Poggianti} {et~al.}(2019){Poggianti}, {Gullieuszik}, {Tonnesen}, {Moretti}, {Vulcani}, {Radovich}, {Jaff{\'e}}, {Fritz}, {Bettoni}, {Franchetto}, {Fasano}, {Bellhouse}, \& {Omizzolo}}]{Poggianti:2019a}
{Poggianti}, B.~M., {Gullieuszik}, M., {Tonnesen}, S., {et~al.} 2019, \mnras, 482, 4466

\bibitem[{{Poggianti} {et~al.}(2017{\natexlab{a}}){Poggianti}, {Jaff{\'e}}, {Moretti}, {Gullieuszik}, {Radovich}, {Tonnesen}, {Fritz}, {Bettoni}, {Vulcani}, {Fasano}, {Bellhouse}, {Hau}, \& {Omizzolo}}]{Poggianti:2017b}
{Poggianti}, B.~M., {Jaff{\'e}}, Y.~L., {Moretti}, A., {et~al.} 2017{\natexlab{a}}, \nat, 548, 304

\bibitem[{{Poggianti} {et~al.}(2017{\natexlab{b}}){Poggianti}, {Moretti}, {Gullieuszik}, {Fritz}, {Jaff{\'e}}, {Bettoni}, {Fasano}, {Bellhouse}, {Hau}, {Vulcani}, {Biviano}, {Omizzolo}, {Paccagnella}, {D'Onofrio}, {Cava}, {Sheen}, {Couch}, \& {Owers}}]{Poggianti:2017a}
{Poggianti}, B.~M., {Moretti}, A., {Gullieuszik}, M., {et~al.} 2017{\natexlab{b}}, \apj, 844, 48

\bibitem[{{Ramatsoku} {et~al.}(2020){Ramatsoku}, {Serra}, {Poggianti}, {Moretti}, {Gullieuszik}, {Bettoni}, {Deb}, {Franchetto}, {van Gorkom}, {Jaff{\'e}}, {Tonnesen}, {Verheijen}, {Vulcani}, {Andati}, {de Blok}, {J{\'o}zsa}, {Kamphuis}, {Kleiner}, {Maccagni}, {Makhathini}, {Moln{\'a}r}, {Ramaila}, {Smirnov}, \& {Thorat}}]{Ramatsoku:2020a}
{Ramatsoku}, M., {Serra}, P., {Poggianti}, B.~M., {et~al.} 2020, \aap, 640, A22

\bibitem[{{Ramatsoku} {et~al.}(2019){Ramatsoku}, {Serra}, {Poggianti}, {Moretti}, {Gullieuszik}, {Bettoni}, {Deb}, {Fritz}, {van Gorkom}, {Jaff{\'e}}, {Tonnesen}, {Verheijen}, {Vulcani}, {Hugo}, {J{\'o}zsa}, {Maccagni}, {Makhathini}, {Ramaila}, {Smirnov}, \& {Thorat}}]{Ramatsoku:2019a}
{Ramatsoku}, M., {Serra}, P., {Poggianti}, B.~M., {et~al.} 2019, \mnras, 487, 4580

\bibitem[{{Rasmussen} {et~al.}(2006){Rasmussen}, {Ponman}, {Mulchaey}, {Miles}, \& {Raychaudhury}}]{Rasmussen:2006a}
{Rasmussen}, J., {Ponman}, T.~J., {Mulchaey}, J.~S., {Miles}, T.~A., \& {Raychaudhury}, S. 2006, \mnras, 373, 653

\bibitem[{{Rhee} {et~al.}(2017){Rhee}, {Smith}, {Choi}, {Yi}, {Jaff{\'e}}, {Candlish}, \& {S{\'a}nchez-J{\'a}nssen}}]{Rhee:2017a}
{Rhee}, J., {Smith}, R., {Choi}, H., {et~al.} 2017, \apj, 843, 128

\bibitem[{{Ricarte} {et~al.}(2020){Ricarte}, {Tremmel}, {Natarajan}, \& {Quinn}}]{Ricarte:2020a}
{Ricarte}, A., {Tremmel}, M., {Natarajan}, P., \& {Quinn}, T. 2020, \apjl, 895, L8

\bibitem[{{Roberts} {et~al.}(2022{\natexlab{a}}){Roberts}, {Lang}, {Trotsenko}, {Bemis}, {Ellison}, {Lin}, {Pan}, {Ignesti}, {Leslie}, \& {van Weeren}}]{Roberts:2022b}
{Roberts}, I.~D., {Lang}, M., {Trotsenko}, D., {et~al.} 2022{\natexlab{a}}, \apj, 941, 77

\bibitem[{{Roberts} \& {Parker}(2020)}]{Roberts:2020a}
{Roberts}, I.~D. \& {Parker}, L.~C. 2020, \mnras, 495, 554

\bibitem[{{Roberts} {et~al.}(2022{\natexlab{b}}){Roberts}, {Parker}, {Gwyn}, {Hudson}, {Carlberg}, {McConnachie}, {Cuillandre}, {Chambers}, {Duc}, {Furusawa}, {Gavazzi}, {Hill}, {Huber}, {Ibata}, {Kilbinger}, {Mei}, {Mellier}, {Miyazaki}, {Oguri}, \& {Wainscoat}}]{Roberts:2022a}
{Roberts}, I.~D., {Parker}, L.~C., {Gwyn}, S., {et~al.} 2022{\natexlab{b}}, \mnras, 509, 1342

\bibitem[{{Roberts} {et~al.}(2021{\natexlab{a}}){Roberts}, {van Weeren}, {McGee}, {Botteon}, {Drabent}, {Ignesti}, {Rottgering}, {Shimwell}, \& {Tasse}}]{Roberts:2021a}
{Roberts}, I.~D., {van Weeren}, R.~J., {McGee}, S.~L., {et~al.} 2021{\natexlab{a}}, \aap, 650, A111

\bibitem[{{Roberts} {et~al.}(2021{\natexlab{b}}){Roberts}, {van Weeren}, {McGee}, {Botteon}, {Ignesti}, \& {Rottgering}}]{Roberts:2021b}
{Roberts}, I.~D., {van Weeren}, R.~J., {McGee}, S.~L., {et~al.} 2021{\natexlab{b}}, \aap, 652, A153

\bibitem[{{Roman-Oliveira} {et~al.}(2021){Roman-Oliveira}, {Chies-Santos}, {Ferrari}, {Lucatelli}, \& {Rodr{\'\i}guez Del Pino}}]{Roman-Oliveira:2021a}
{Roman-Oliveira}, F., {Chies-Santos}, A.~L., {Ferrari}, F., {Lucatelli}, G., \& {Rodr{\'\i}guez Del Pino}, B. 2021, \mnras, 500, 40

\bibitem[{{S{\'a}nchez-Garc{\'\i}a} {et~al.}(2023){S{\'a}nchez-Garc{\'\i}a}, {Cervantes Sodi}, {Fritz}, {Moretti}, {Poggianti}, {George}, {Gullieuszik}, {Vulcani}, {Fasano}, \& {Tawfeek}}]{Sanchez-Garcia:2023a}
{S{\'a}nchez-Garc{\'\i}a}, O., {Cervantes Sodi}, B., {Fritz}, J., {et~al.} 2023, \apj, 945, 99

\bibitem[{{Serra} {et~al.}(2023){Serra}, {Maccagni}, {Kleiner}, {Moln{\'a}r}, {Ramatsoku}, {Loni}, {Loi}, {de Blok}, {Bryan}, {Dettmar}, {Frank}, {van Gorkom}, {Govoni}, {Iodice}, {J{\'o}zsa}, {Kamphuis}, {Kraan-Korteweg}, {Loubser}, {Murgia}, {Oosterloo}, {Peletier}, {Pisano}, {Smith}, {Trager}, \& {Verheijen}}]{Serra:2023a}
{Serra}, P., {Maccagni}, F.~M., {Kleiner}, D., {et~al.} 2023, \aap, 673, A146

\bibitem[{{Simmons} {et~al.}(2014){Simmons}, {Melvin}, {Lintott}, {Masters}, {Willett}, {Keel}, {Smethurst}, {Cheung}, {Nichol}, {Schawinski}, {Rutkowski}, {Kartaltepe}, {Bell}, {Casteels}, {Conselice}, {Almaini}, {Ferguson}, {Fortson}, {Hartley}, {Kocevski}, {Koekemoer}, {McIntosh}, {Mortlock}, {Newman}, {Ownsworth}, {Bamford}, {Dahlen}, {Faber}, {Finkelstein}, {Fontana}, {Galametz}, {Grogin}, {Gr{\"u}tzbauch}, {Guo}, {H{\"a}u{\ss}ler}, {Jek}, {Kaviraj}, {Lucas}, {Peth}, {Salvato}, {Wiklind}, \& {Wuyts}}]{Simmons:2014a}
{Simmons}, B.~D., {Melvin}, T., {Lintott}, C., {et~al.} 2014, \mnras, 445, 3466

\bibitem[{{Skibba} {et~al.}(2012){Skibba}, {Masters}, {Nichol}, {Zehavi}, {Hoyle}, {Edmondson}, {Bamford}, {Cardamone}, {Keel}, {Lintott}, \& {Schawinski}}]{Skibba:2012a}
{Skibba}, R.~A., {Masters}, K.~L., {Nichol}, R.~C., {et~al.} 2012, \mnras, 423, 1485

\bibitem[{{Smethurst} {et~al.}(2017){Smethurst}, {Lintott}, {Bamford}, {Hart}, {Kruk}, {Masters}, {Nichol}, \& {Simmons}}]{Smethurst:2017a}
{Smethurst}, R.~J., {Lintott}, C.~J., {Bamford}, S.~P., {et~al.} 2017, \mnras, 469, 3670

\bibitem[{{Smith} {et~al.}(2022){Smith}, {Shinn}, {Tonnesen}, {Calder{\'o}n-Castillo}, {Crossett}, {Jaffe}, {Roberts}, {McGee}, {George}, {Vulcani}, {Gullieuszik}, {Moretti}, {Poggianti}, \& {Shin}}]{Smith:2022a}
{Smith}, R., {Shinn}, J.-H., {Tonnesen}, S., {et~al.} 2022, \apj, 934, 86

\bibitem[{{Smith} {et~al.}(2010){Smith}, {Lucey}, {Hammer}, {Hornschemeier}, {Carter}, {Hudson}, {Marzke}, {Mouhcine}, {Eftekharzadeh}, {James}, {Khosroshahi}, {Kourkchi}, \& {Karick}}]{Smith:2010a}
{Smith}, R.~J., {Lucey}, J.~R., {Hammer}, D., {et~al.} 2010, \mnras, 408, 1417

\bibitem[{{Sun} {et~al.}(2007){Sun}, {Donahue}, \& {Voit}}]{Sun:2007a}
{Sun}, M., {Donahue}, M., \& {Voit}, G.~M. 2007, \apj, 671, 190

\bibitem[{{Tanaka} {et~al.}(2023){Tanaka}, {Koike}, {Naito}, {Shibata}, {Usuda-Sato}, {Yamaoka}, {Ando}, {Ito}, {Kobayashi}, {Kofuji}, {Kuwata}, {Nakano}, {Shimakawa}, {Tadaki}, {Takebayashi}, {Tsuchiya}, {Umemoto}, \& {Bottrell}}]{Tanaka:2023a}
{Tanaka}, M., {Koike}, M., {Naito}, S., {et~al.} 2023, \pasj, 75, 986

\bibitem[{{Tempel} {et~al.}(2014){Tempel}, {Tamm}, {Gramann}, {Tuvikene}, {Liivam{\"a}gi}, {Suhhonenko}, {Kipper}, {Einasto}, \& {Saar}}]{Tempel:2014a}
{Tempel}, E., {Tamm}, A., {Gramann}, M., {et~al.} 2014, \aap, 566, A1

\bibitem[{{Troncoso-Iribarren} {et~al.}(2020){Troncoso-Iribarren}, {Padilla}, {Santander}, {Lagos}, {Garc{\'\i}a-Lambas}, {Rodr{\'\i}guez}, \& {Contreras}}]{Troncoso-Iribarren:2020a}
{Troncoso-Iribarren}, P., {Padilla}, N., {Santander}, C., {et~al.} 2020, \mnras, 497, 4145

\bibitem[{{Villanueva} {et~al.}(2022){Villanueva}, {Bolatto}, {Vogel}, {Brown}, {Wilson}, {Zabel}, {Ellison}, {Stevens}, {Jim{\'e}nez Donaire}, {Spekkens}, {Tharp}, {Davis}, {Parker}, {Roberts}, {Basra}, {Boselli}, {Catinella}, {Chung}, {Cortese}, {Lee}, \& {Watts}}]{Villanueva:2022a}
{Villanueva}, V., {Bolatto}, A.~D., {Vogel}, S., {et~al.} 2022, \apj, 940, 176

\bibitem[{{Vogt} {et~al.}(2004){Vogt}, {Haynes}, {Giovanelli}, \& {Herter}}]{Vogt:2004a}
{Vogt}, N.~P., {Haynes}, M.~P., {Giovanelli}, R., \& {Herter}, T. 2004, \aj, 127, 3300

\bibitem[{{Vollmer} {et~al.}(2004){Vollmer}, {Balkowski}, {Cayatte}, {van Driel}, \& {Huchtmeier}}]{Vollmer:2004a}
{Vollmer}, B., {Balkowski}, C., {Cayatte}, V., {van Driel}, W., \& {Huchtmeier}, W. 2004, \aap, 419, 35

\bibitem[{{Vollmer} {et~al.}(2001){Vollmer}, {Cayatte}, {Balkowski}, \& {Duschl}}]{Vollmer:2001a}
{Vollmer}, B., {Cayatte}, V., {Balkowski}, C., \& {Duschl}, W.~J. 2001, \apj, 561, 708

\bibitem[{{Vulcani} {et~al.}(2020){Vulcani}, {Fritz}, {Poggianti}, {Bettoni}, {Franchetto}, {Moretti}, {Gullieuszik}, {Jaff{\'e}}, {Biviano}, {Radovich}, \& {Mingozzi}}]{Vulcani:2020a}
{Vulcani}, B., {Fritz}, J., {Poggianti}, B.~M., {et~al.} 2020, \apj, 892, 146

\bibitem[{{Vulcani} {et~al.}(2018{\natexlab{a}}){Vulcani}, {Poggianti}, {Gullieuszik}, {Moretti}, {Tonnesen}, {Jaff{\'e}}, {Fritz}, {Fasano}, \& {Bettoni}}]{Vulcani:2018b}
{Vulcani}, B., {Poggianti}, B.~M., {Gullieuszik}, M., {et~al.} 2018{\natexlab{a}}, \apjl, 866, L25

\bibitem[{{Vulcani} {et~al.}(2018{\natexlab{b}}){Vulcani}, {Poggianti}, {Jaff{\'e}}, {Moretti}, {Fritz}, {Gullieuszik}, {Bettoni}, {Fasano}, {Tonnesen}, \& {McGee}}]{Vulcani:2018a}
{Vulcani}, B., {Poggianti}, B.~M., {Jaff{\'e}}, Y.~L., {et~al.} 2018{\natexlab{b}}, \mnras, 480, 3152

\bibitem[{{Vulcani} {et~al.}(2021){Vulcani}, {Poggianti}, {Moretti}, {Franchetto}, {Bacchini}, {McGee}, {Jaff{\'e}}, {Mingozzi}, {Werle}, {Tomi{\v{c}}i{\'c}}, {Fritz}, {Bettoni}, {Wolter}, \& {Gullieuszik}}]{Vulcani:2021a}
{Vulcani}, B., {Poggianti}, B.~M., {Moretti}, A., {et~al.} 2021, \apj, 914, 27

\bibitem[{{Vulcani} {et~al.}(2022){Vulcani}, {Poggianti}, {Smith}, {Moretti}, {Jaff{\'e}}, {Gullieuszik}, {Fritz}, \& {Bellhouse}}]{Vulcani:2022a}
{Vulcani}, B., {Poggianti}, B.~M., {Smith}, R., {et~al.} 2022, \apj, 927, 91

\bibitem[{{Walmsley} {et~al.}(2023){Walmsley}, {G{\'e}ron}, {Kruk}, {Scaife}, {Lintott}, {Masters}, {Dawson}, {Dickinson}, {Fortson}, {Garland}, {Mantha}, {O'Ryan}, {Popp}, {Simmons}, {Baeten}, \& {Macmillan}}]{Walmsley:2023a}
{Walmsley}, M., {G{\'e}ron}, T., {Kruk}, S., {et~al.} 2023, \mnras, 526, 4768

\bibitem[{{Walmsley} {et~al.}(2022){Walmsley}, {Scaife}, {Lintott}, {Lochner}, {Etsebeth}, {G{\'e}ron}, {Dickinson}, {Fortson}, {Kruk}, {Masters}, {Mantha}, \& {Simmons}}]{Walmsley:2022a}
{Walmsley}, M., {Scaife}, A. M.~M., {Lintott}, C., {et~al.} 2022, \mnras, 513, 1581

\bibitem[{{Walmsley} {et~al.}(2020){Walmsley}, {Smith}, {Lintott}, {Gal}, {Bamford}, {Dickinson}, {Fortson}, {Kruk}, {Masters}, {Scarlata}, {Simmons}, {Smethurst}, \& {Wright}}]{Walmsley:2020a}
{Walmsley}, M., {Smith}, L., {Lintott}, C., {et~al.} 2020, \mnras, 491, 1554

\bibitem[{{Watts} {et~al.}(2023){Watts}, {Cortese}, {Catinella}, {Brown}, {Wilson}, {Zabel}, {Roberts}, {Davis}, {Thorp}, {Chung}, {Stevens}, {Ellison}, {Spekkens}, {Parker}, {Bah{\'e}}, {Villanueva}, {Jim{\'e}nez-Donaire}, {Bisaria}, {Boselli}, {Bolatto}, \& {Lee}}]{Watts:2023a}
{Watts}, A.~B., {Cortese}, L., {Catinella}, B., {et~al.} 2023, \pasa, 40, e017

\bibitem[{{Willett} {et~al.}(2013){Willett}, {Lintott}, {Bamford}, {Masters}, {Simmons}, {Casteels}, {Edmondson}, {Fortson}, {Kaviraj}, {Keel}, {Melvin}, {Nichol}, {Raddick}, {Schawinski}, {Simpson}, {Skibba}, {Smith}, \& {Thomas}}]{Willett:2013a}
{Willett}, K.~W., {Lintott}, C.~J., {Bamford}, S.~P., {et~al.} 2013, \mnras, 435, 2835

\bibitem[{{Yagi} {et~al.}(2007){Yagi}, {Komiyama}, {Yoshida}, {Furusawa}, {Kashikawa}, {Koyama}, \& {Okamura}}]{Yagi:2007a}
{Yagi}, M., {Komiyama}, Y., {Yoshida}, M., {et~al.} 2007, \apj, 660, 1209

\bibitem[{{Yagi} {et~al.}(2010){Yagi}, {Yoshida}, {Komiyama}, {Kashikawa}, {Furusawa}, {Okamura}, {Graham}, {Miller}, {Carter}, {Mobasher}, \& {Jogee}}]{Yagi:2010a}
{Yagi}, M., {Yoshida}, M., {Komiyama}, Y., {et~al.} 2010, \aj, 140, 1814

\bibitem[{{Zinger} {et~al.}(2024){Zinger}, {Joshi}, {Pillepich}, {Rohr}, \& {Nelson}}]{Zinger:2024a}
{Zinger}, E., {Joshi}, G.~D., {Pillepich}, A., {Rohr}, E., \& {Nelson}, D. 2024, \mnras, 527, 8257

\end{thebibliography}

\begin{appendix}
\onecolumn

\section{Additional Material}

\begin{figure*}
    \begin{centering}
	\includegraphics[width=17cm]{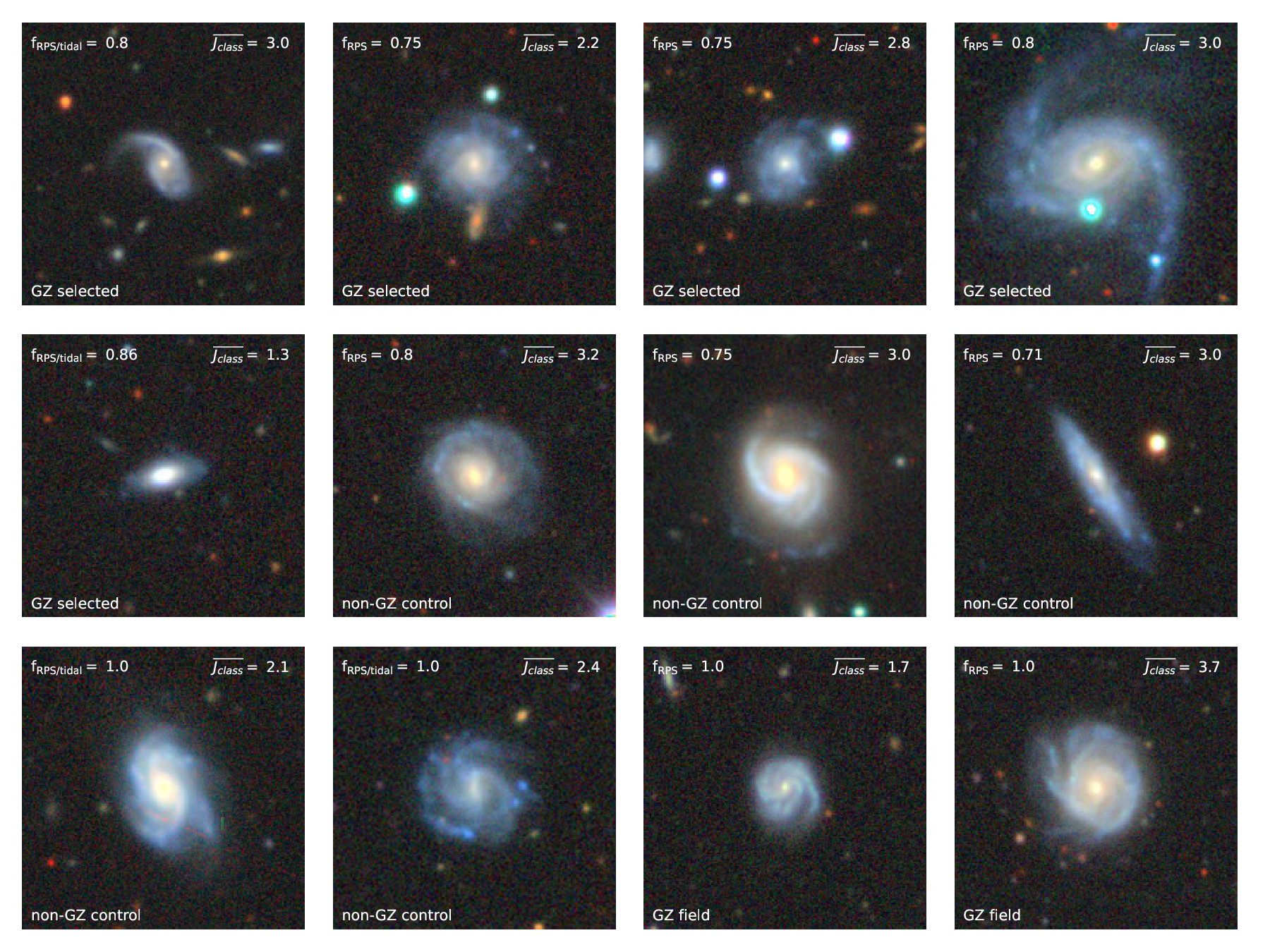}
    \caption{Legacy Survey 3 colour images of several additional example expert classified RPS candidates found in this study, similar to Fig. \ref{fig:new_cand_images}.}
    \label{fig:extra_cand_images}
    \end{centering}
\end{figure*}

\begin{table}
\caption{Names of galaxy samples used in this study}
\begin{tabular}{p{1in}p{0.5in}p{0.5in}p{2.5in}p{2in}}
\hline
Sample name & No.\ of galaxies & Section & Description & Other notes \\ \hline

Known RPS candidates & 212 & Sect.\ \ref{sec:morphologies} & A compilation of known ram pressure stripping candidates that overlap with Galaxy Zoo 2 data & Sources given in Table \ref{tab:GZ_jf_sources}. A subset of 100 of these are used in Sect. \ref{sec:GZ_motivated}\\

Known strong RPS candidates & 58 & Sect.\ \ref{sec:strong_morphologies} & Known RPS candidates that only have optical features and a high RPS `strength' & Includes known RPS sample galaxies classified in Sect. \ref{sec:samp} for this study\\

Known Radio RPS candidates & 134 & Sect.\ \ref{sec:radio_only} & Known RPS candidates that have radio features & Includes galaxies that have optical features\\
&&&& \\

Known radio only RPS candidates & 75 & Sect.\ \ref{sec:radio_only} & Known RPS candidates that only have radio features & No optical features reported in other studies or from classification in Sect. \ref{sec:samp} \\

Field comparison sample & 1696 & Sect.\ \ref{sec:comparisons} & A sample of galaxies that are magnitude and redshift matched to the known RPS sample that are away from clusters & Cut to match strong and radio-identified RPS samples where appropriate\\

Cluster comparison sample &  1696 & Sect.\ \ref{sec:comparisons} & A sample of galaxies that are magnitude and redshift matched to the known RPS sample that are around cluster cores & Cut to match strong and radio-identified RPS samples where appropriate\\

GZ-selected sample & 319 & Sect.\ \ref{sec:motivated} & GZ2 classified disk galaxies that have properties consistent with the known RPS sample in clusters& \\

Non-GZ control sample & 319 & Sect.\ \ref{sec:controls} & A sample of disk galaxies around cluster cores that are magnitude and redshift matched to the GZ-selected sample, but do not have unwinding and odd morphologies consistent with the GZ selected sample. & \\

GZ-field sample & 319 & Sect.\ \ref{sec:false} & GZ2 classified disk galaxies that have properties consistent with the known RPS sample, but are not in clusters.& \\

Expert classified RPS candidates & 101 & Sect.\ \ref{sec:new_cands} & A combined subsample of the GZ-selected sample and the non-GZ control sample that  have been classified by experts as being ram pressure stripped &  \\
\hline
\end{tabular}
\label{tab:samp_list}
\end{table}

\begin{table*}
	\centering
	\caption{Questionnaire given to expert classifiers for identifying ram pressure stripping features.}
	\label{tab:GASP_questions}
	\begin{tabular}{|p{2in}|p{4.7in}|} 
		\hline
		Question/task & Response \\
		\hline
Do you see any unusual features in this galaxy?
& -- Merger/Tidal interaction - It looks like the galaxy is merging or has been affected by gravitational interactions\\
& -- Ram Pressure - The galaxy has tails or other features (such as unwinding spiral arms) indicative of ram pressure stripping\\
& -- Unsure - It appears that something is disturbing the galaxy, but I can't tell if it is ram pressure stripping or something else\\
& -- Ram Pressure + Tidal - Both ram pressure stripping + tidal interaction could be at play\\
& -- Something Else/Nothing - This galaxy doesn't appear to have any unusual features, or there are features\\
& \hspace{0.2cm} which are not described above\\
		\hline
$^{\dagger}$What ram pressure stripping features do you see? (select all that apply)
& -- Tails (faint diffuse extra-planar material and/or extra-planar knots of star formation) \\
& -- Star formation on one side of the disk indicative of gas compression in the leading edge / shock front (may be C-shaped)\\
& -- Spiral-arm asymmetries or deformations \\
& -- Something else (please note this in the comment section) \\
		\hline
$^{\dagger}$On a scale of 1 to 5, how strong would you describe the strength of the ram pressure stripping?
& -- 5 - This galaxy contains very clear and obvious ram pressure signatures (spectacular cases of jellyfish)\\
& -- 4 - This galaxy contains clear ram pressure features\\
& -- 3 - This galaxy contains probable ram pressure features that are visible in the image\\
& -- 2 - This galaxy might contain some features consistent with ram pressure stripping, but it is hard to be certain\\
& -- 1 - There is something unusual about this galaxy, which may be consistent with ram pressure stripping, but it is not very clear\\
& -- 0 - On closer inspection, maybe I don't see ram pressure features any more\\
		\hline
$^{\dagger}$Can you estimate the tail direction, or the direction of the ram pressure wind?
& -- Yes, I can give an estimate of the tail direction\\
& -- No, I can't tell which direction the ram pressure is acting\\
		\hline
$^{\dagger \ddagger}$Draw in the direction of the tail/ram pressure stripping (i.e. opposite to the projected direction of motion of the galaxy). Start your line in the centre of the image.
& -- (Classifier draws in the tail direction or direction of the ram pressure wind)\\
		\hline
If there is anything interesting about this galaxy you wish to note, write it below
& -- User open for comments\\
		\hline
	\end{tabular}
    \tablefoot{
\tablefoottext{${\dagger}$}{requires the classifier to answer question 1 with Ram Pressure, Unsure, or Ram Pressure + Tidal to receive this question.}
\tablefoottext{${\ddagger}$}{requires the classifier to answer question 4 with Yes to receive this question.}
}
\end{table*}

\label{lastpage}
\end{appendix}

\end{document}